\documentclass[preprint,11.5pt]{elsarticle}

\usepackage{amssymb}
\usepackage{algorithmic}
\usepackage{algorithm}
\usepackage{multirow}
\usepackage{makecell}
\usepackage{array}
\usepackage{amsmath,amsfonts}
\usepackage{pifont}
\usepackage{booktabs}
\usepackage{color}
\usepackage{enumitem}
\usepackage{hyperref}
\usepackage{setspace}
\usepackage{subfigure}
\usepackage{geometry}
\usepackage{tabularx}
\usepackage{amsfonts}

\journal{Omega}

\begin{document}	
\begin{frontmatter}
		\begin{spacing}{2.0}

		\title{Towards Popularity-Aware Recommendation: A Multi-Behavior Enhanced Framework with Orthogonality Constraint}
			
		\author[inst1]{Yishan Han\corref{correspondingauthor}}
            \cortext[correspondingauthor]{Corresponding author at: School of Management, Xi'an Jiaotong University, No.28, West Xianning Road, Xi'an, Shaanxi, 710049, P.R. China.}
		\ead{13.yishan.han@gmail.com}			
		\address[inst1]{{Center for Intelligent Decision-Making and Machine Learning, School of Management}, {Xi'an Jiaotong University}, 
				{Xi'an}, 
                {Shaanxi},
				{P.R.China}}
		\address[inst2]{{Department of Marketing, School of Management}, {Xi'an Jiaotong University}, 
				{Xi'an},
                {Shaanxi},
				{P.R.China}}
		\author[inst1]{Biao Xu}
		\ead{xubiao.xjtu@gmail.com }  
  \author[inst1]{Yao Wang}
		\ead{yao.s.wang@gmail.com}
		\author[inst2]{Shanxing Gao}
        \ead{gaozn@mail.xjtu.edu.cn}

\begin{abstract}
\indent Top-$K$ recommendation involves inferring latent user preferences and generating personalized recommendations accordingly, which is now ubiquitous in various decision systems. 
Nonetheless, recommender systems usually suffer from severe \textit{popularity bias}, leading to the over-recommendation of popular items. 
Such a bias deviates from the central aim of reflecting user preference faithfully, compromising both customer satisfaction and retailer profits.
Despite the prevalence, existing methods tackling popularity bias still have limitations due to the considerable accuracy-debias tradeoff and the sensitivity to extensive parameter selection, further exacerbated by the extreme sparsity in positive user-item interactions.

In this paper, we present a \textbf{Pop}ularity-aware top-$K$ recommendation algorithm integrating multi-behavior \textbf{S}ide \textbf{I}nformation (PopSI), aiming to enhance recommendation accuracy and debias performance simultaneously.
Specifically, by leveraging multiple user feedback that mirrors similar user preferences and formulating it as a three-dimensional tensor, PopSI can utilize all slices to capture the desiring user preferences effectively. 
Subsequently, we introduced a novel orthogonality constraint to refine the estimated item feature space, enforcing it to be invariant to item popularity features thereby addressing our model's sensitivity to popularity bias.
Comprehensive experiments on real-world e-commerce datasets demonstrate the general improvements of PopSI over state-of-the-art debias methods with a marginal accuracy-debias tradeoff and scalability to practical applications.
The source code for our algorithm and experiments is available at \url{https://github.com/Eason-sys/PopSI}.
\end{abstract}

\end{spacing}
\end{frontmatter}
 
\section{Introduction}\label{sec_intro}
\begin{spacing}{2.0}
\vspace{-2mm}
In this era of information overabundance, recommender systems are becoming increasingly important decision aids for consumers, as well as an integral part of the business models for many organizations and platforms. Recommendation algorithm typically works by analyzing patterns of user behaviors to recommend items that best fit users' personal preferences, which finds a wide range of applications across diverse domains, including but not limited to e-commerce businesses \cite{chu2020position}, online job markets \cite{kokkodis2023good}, and music streaming services \cite{farias2019learning}.
In this paper, we focus on top-$K$ recommendations with implicit feedback in e-commerce. Unlike explicit rating records, implicit feedback involves user behaviors such as click, collect, and purchase, which are often more readily available and prevalent in practical scenarios. The objective of implicit top-$K$ recommendation is to infer the hidden user preferences and generate personalized size-$K$ lists of recommendations accordingly \cite{rendle2012bpr, he2018pseudo, adomavicius2016classification}. 
	
    As has been extensively demonstrated both empirically and theoretically, top-$K$ recommendation algorithms often suffer from severe \textit{popularity bias}, a phenomenon where popular items are overly recommended than their popularity would warrant \cite{steck2011item, abdollahpouri2019managing, chen2023bias}. 
    While item popularity doesn't necessarily pose negative implications \cite{zhao2022popularity, zhang2021causal}, a too strong focus on popular items can be disadvantageous both for users and providers. 
    Users might perceive the recommendations as predictable, thereby failing to fulfill their desire for discovery. Additionally, low-popularity items hold significance as they can offer serendipitous and novel experiences, essential for broadening users' interests.
    Providers, on the other hand, not only fail to supply adequate discovery support but also miss the potential profits from the niche items, ultimately risking decreased user engagement. \cite{zhu2021popularity}.
	
\begin{figure}[t]\vspace{0mm}
		\centering
		\includegraphics[width=0.66\linewidth]{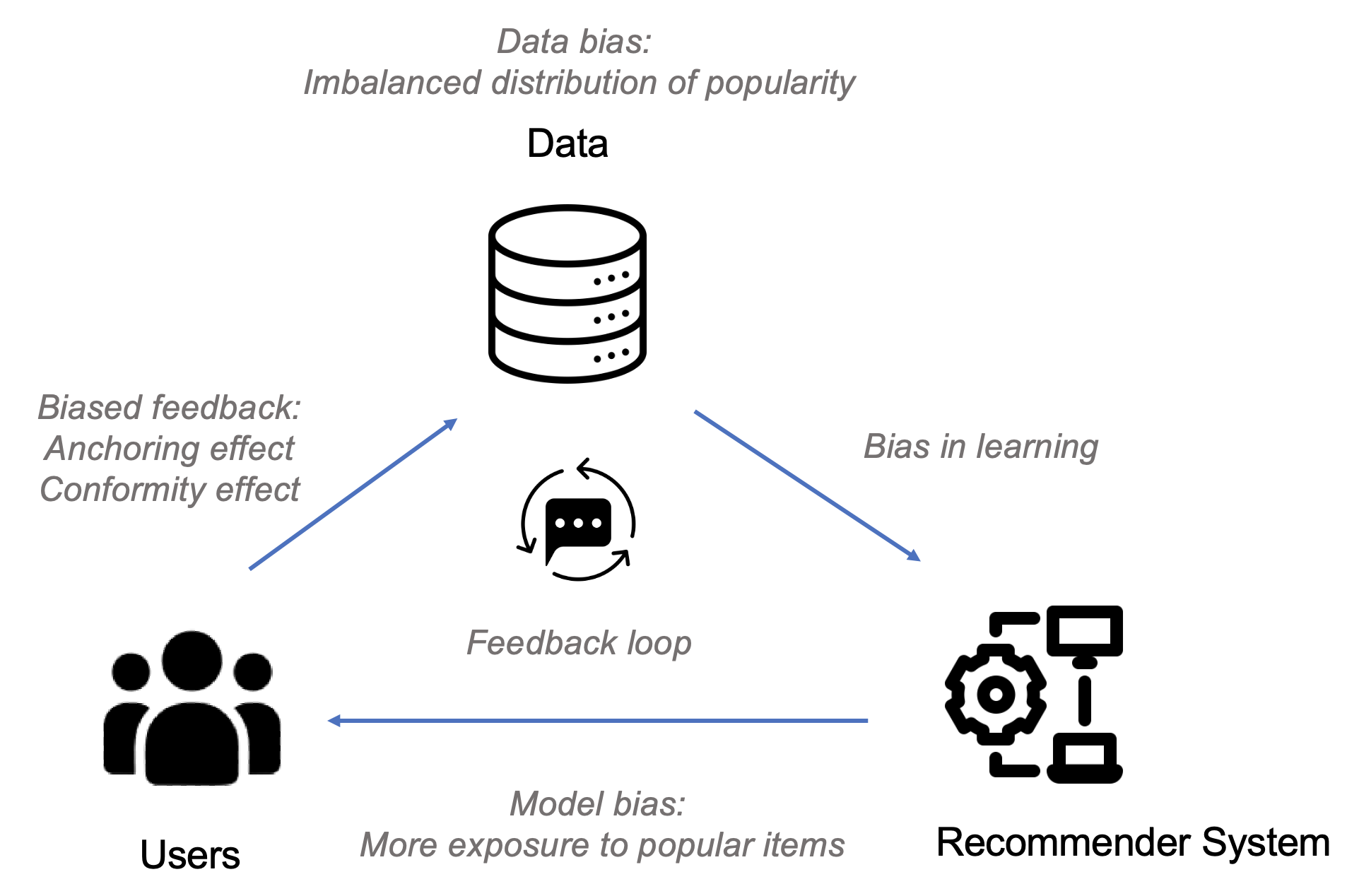}
		\caption{The sources of popularity bias and the recommendation feedback loop.}\label{loop}
	\end{figure}

    The origin of popularity bias primarily stems from two distinct sources \cite{steck2011item, abdollahpouri2019managing, chen2023bias}. From the data side, the collected user-item feedback often shows \textit{long-tail} distribution in item frequency with most interactions focused on a small number of popular items (data bias) \cite{brynjolfsson2006niches, Oestreicher2012Recommendation}. From the model side, the recommender system, trained on such imbalanced data, often produces higher preference scores for popular items even among items equally liked by a user, resulting in over-recommending popular items (model bias) \cite{zhu2021popularity, abdollahpouri2020connection}. In the long run, popularity bias could accumulate through the recommendation feedback loop, causing undesired effects such as filter bubbles and echo chambers \cite{ aridor2020deconstructing, mansoury2020feedback}, alongside amplifying the anchoring effects \cite{adomavicius2013recommender}  and the conformity effect \cite{canamares2018should}, as shown in Figure~\ref{loop}.

    Given the high practical importance of this problem, an increasing number of studies have been done to mitigate the harmful influence of popularity bias. Representative methodologies include inverse propensity scoring (IPS) \cite{steck2011item, schnabel2016recommendations}, causal intervention \cite{wei2021model, zheng2021disentangling}, preference re-ranking \cite{abdollahpouri2019managing, zhu2021popularity}, and regularization-based methods \cite{zhu2021popularity, bonner2018causal}. 
        IPS focuses on the data bias and removes popularity bias by assigning weights to the imbalanced training data according to the item popularity.
        Causal intervention methods attempt to model and remove the causal effect that item popularity has on the predicted preference scores. 
        Re-ranking methods apply post-processing score/rank adjustments to the recommendation results produced by off-the-shelf algorithms such as matrix factorization.
        
    Despite their effectiveness, existing methods still exhibit certain limitations. Prevailing debiasing mechanisms often rely on an implicit strategy to model item popularity and its relationship to recommendation results, that is, adjusting the biased preference scores inversely proportional to item popularity. Though straightforward and effective, this approach may undermine the utility of other latent item features thus leading to a considerable accuracy-debias tradeoff. Additionally, the extreme scarcity of positive user-item interactions can further exacerbate this issue. Moreover, such methods often require extensive hyper-parameter tuning or additional model assumptions, which hinders their practical applications.
    This prompts the pivotal question: \textit{How can we design an essential popularity-aware algorithm that is both effective and efficient while preserving satisfactory recommendation accuracy?}

\begin{figure}[t]\vspace{-0mm}
    	\centering
    	\subfigure[Number of items a user purchased]
    	{
    		\begin{minipage}[b]{0.40\linewidth}
    			\includegraphics[width=1\linewidth]{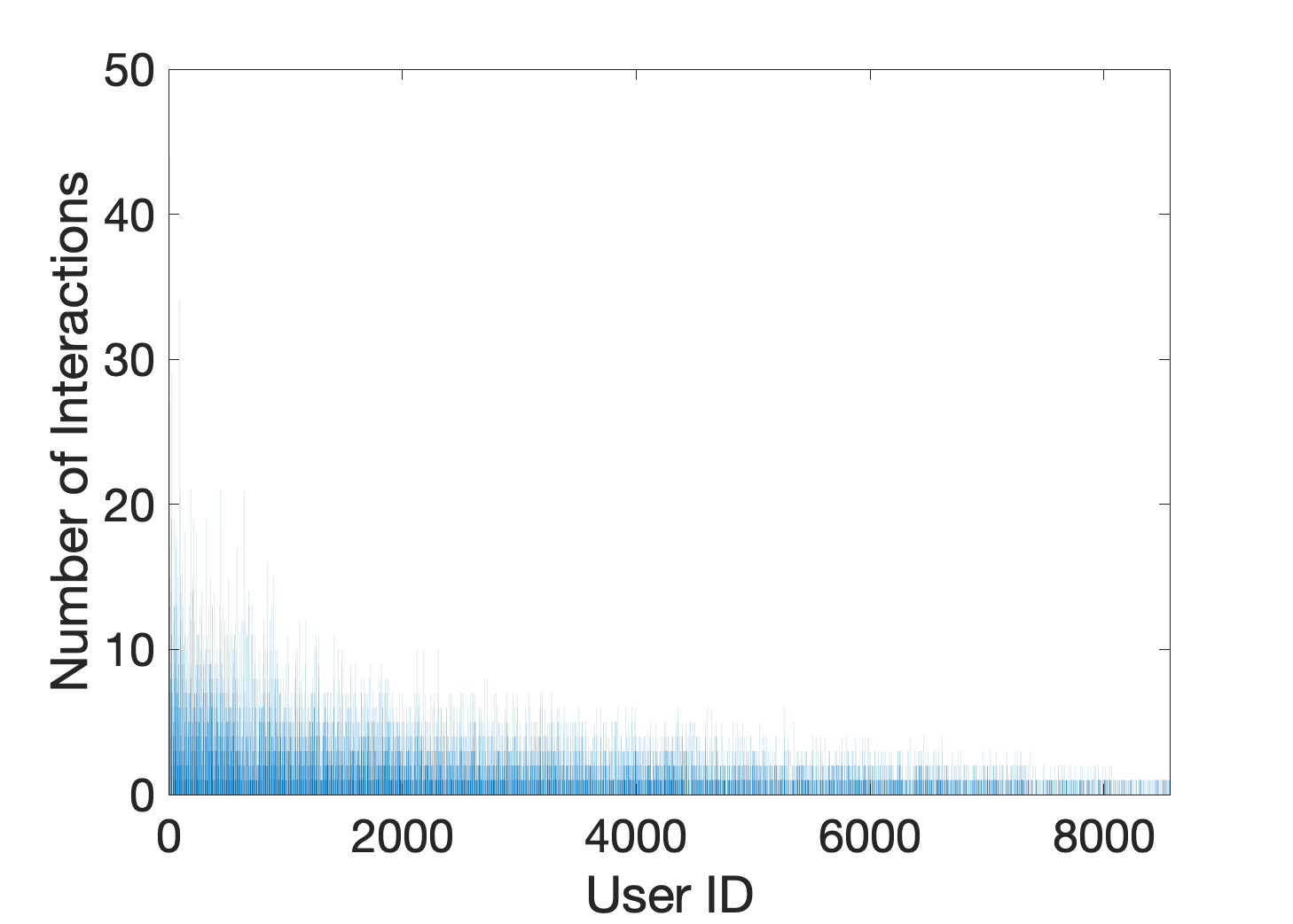}\vspace{-5pt}
    		\end{minipage}
    	}\hspace{-5mm}
    	\subfigure[Number of users purchased an item ]
    	{
    		\begin{minipage}[b]{0.40\linewidth}
    			\includegraphics[width=1\linewidth]{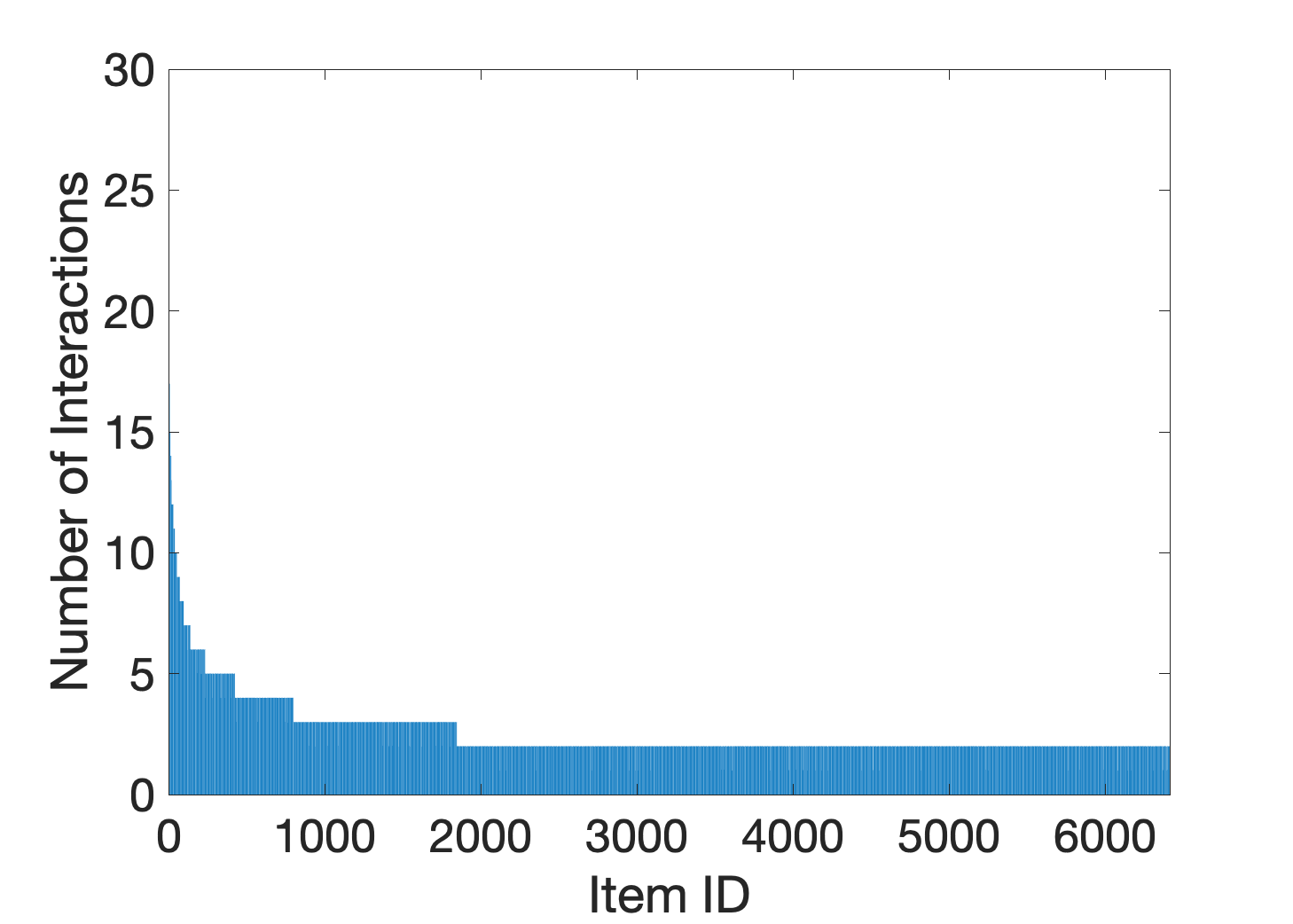}\vspace{-5pt}
    		\end{minipage}
    	}\hspace{-5mm}
    	\caption{The sparsity of a sales transaction record subset from the Tmall dataset.}\label{sparse}
    \end{figure}

    To this end, it is crucial to address two fundamental challenges: data sparsity and intrinsic popularity debias design.
    Implicit feedback, though often more accessible than explicit ratings, also frequently encounters a notable sparsity within contemporary recommender systems \cite{he2018pseudo}, causing compromised performance and difficulties in downstream tasks like popularity debias design.
    Consider the sales transaction records collected by online e-commerce platforms. While serving as a vital indicator of users' purchase propensities, the purchase records are especially sparse due to the low probability associated with a particular customer purchasing a specific product.
    Figure \ref{sparse} illustrates the sparsity of a relatively dense subset of the transaction records from the Tmall dataset. As can be observed, the number of items a user has purchased is mostly below 5 with many instances 0, while the number of users engaged with an item is mostly below 3. Together, it would be extremely hard to extract meaningful user and item features for accurate purchase preference estimation.
	
\begin{figure}[t]\vspace{0mm}
		\centering
		\includegraphics[width=0.9\linewidth]{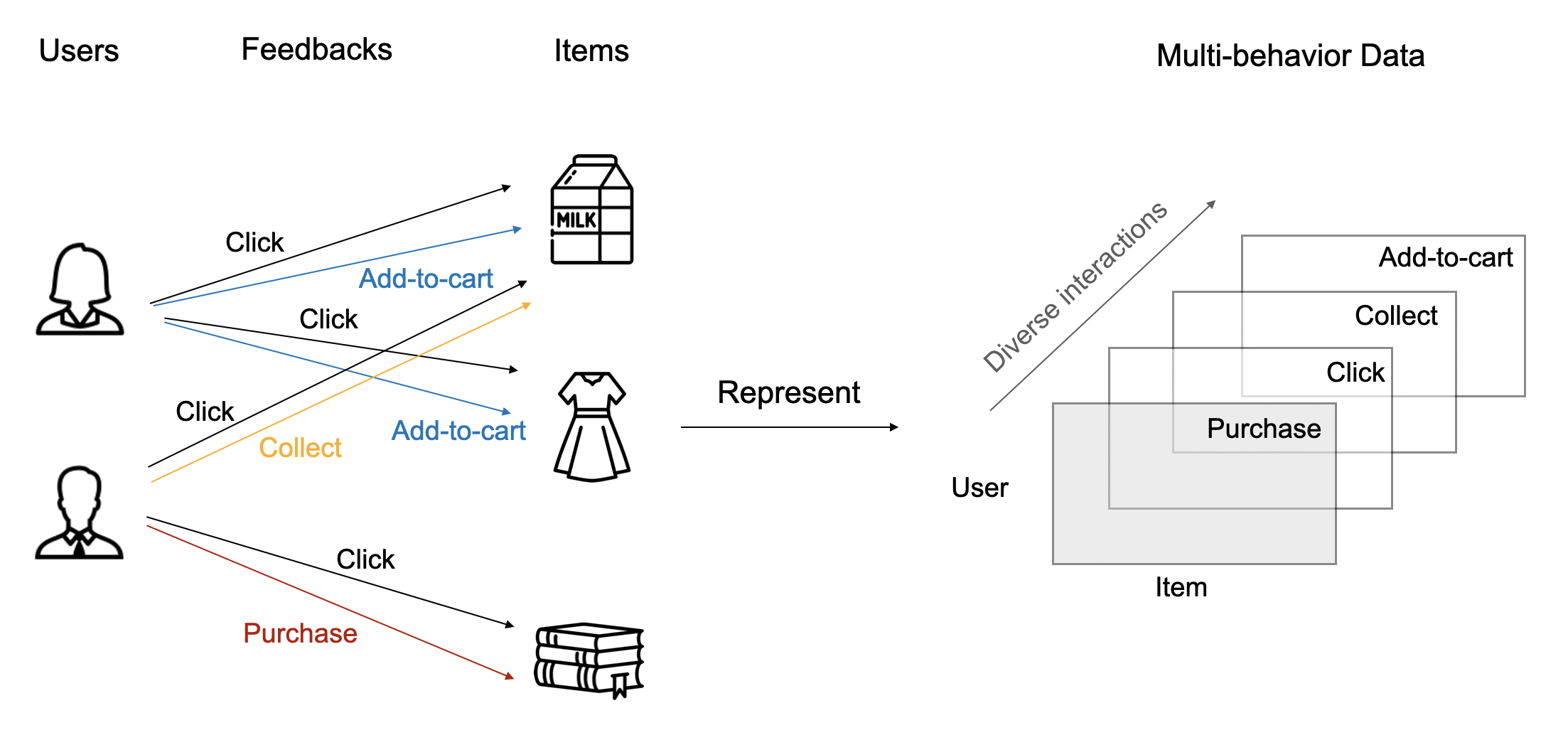}
		\caption{A toy example of multiple feedback between users and items in e-commerce businesses.}\label{multi-behavior}
    \end{figure}

    In light of these limitations, we propose a novel top-$K$ recommendation method integrating multiple user feedback with popularity awareness. To start, we adopt the concept of the latent factor model, which learns to map users and items into corresponding low-rank feature spaces \cite{rendle2012bpr, koren2009matrix}. Then, these dense latent factors enable similarity computations between arbitrary pairs of users and items and thus can address the data sparsity problem. Another strategy countering sparsity is to incorporate auxiliary information, such as social network relationships \cite{zhao2014leveraging}, text reviews on items \cite{duan2022combining, guo2020consumer}, and diverse user preferences \cite{li2021personalized, ren2024consumer}. Beyond primary sales transactions, e-commerce platforms can capture various user-item interactions with different behavioral types, such as click history, collection status, and add-to-cart actions, as shown in Figure~\ref{multi-behavior}. These additional interactions, also referred to as side information, typically exhibit comparatively lower sparsity. Furthermore, the latent user and item features in these heterogeneous behaviors are aligned \cite{farias2019learning}, together providing valuable information for building fine-grained recommendations. \textit{This prompts us to use these less sparse multi-behavior data to assist in estimating the user purchase preference.} 
    Another benefit favoring latent factor models is that such a mechanism facilitates the integration of regularization-based or post-processing debias designs with ease. Previous debiasing methods utilizing such an approach failed to combine multiple user feedback and they mostly choose an implicit way to mitigate popularity bias, which suffers accuracy-debais tradeoff to an undesired extent. \textit{This motivates us to frame an explicit way for modeling item popularity and designing a debias approach without direct score/rank adjustments.}
	
    As stated above, designing an effective and efficient popularity-aware algorithm for top-$K$ recommendation involves considering two crucial elements: adopting an intrinsic modeling approach to employ multiple behaviors for accurate estimation of preferences effectively, and devising an explicit and straightforward debiasing method to mitigate the influence of popularity bias. 
	The main contributions of our paper are summarized as follows:
	
    \textbf{(1) Popularity-aware recommendation design.}  We propose a novel \textbf{Pop}ularity-aware top-$K$ recommendation algorithm with multi-behavior  \textbf{S}ide \textbf{I}nformation  (\textbf{PopSI}). By incorporating diverse user-item interactions and formulating them as a three-dimensional tensor, PopSI proceeds by first utilizing all slices to better characterize the desiring user preference. Subsequently, we explicitly embed popularity as a distinct item feature and refine the estimated item feature space to be invariant to the popularity features by orthogonal projection. This novel and deliberate adjustment addresses the sensitivity of our approach to the influence of popularity bias while preserving the expressiveness of other latent features.
 
    \textbf{(2) Performance enhancements.} We evaluated the effectiveness of our method using real-world e-commerce datasets. Experimental results demonstrate a substantial improvement in recommendation accuracy alongside enhanced debias performance over existing debias approaches with a marginal tradeoff, which is uncommon in other debias methods. Additionally, leveraging multi-behavior data alone also contributes to the debias performance.
		
    \textbf{(3) Practical implications.} The datasets employed in commercial applications are typically extensive and complex, requiring efficient algorithms for rapid processing to provide timely recommendations. Utilizing primarily singular value decomposition on sparse matrices and orthogonal projections, our method exhibits low computational complexity with minimal parameter tuning. Notably, it is easy to implement and scales efficiently to large datasets, thus finding value in product promotion for various decision support systems.

    The remainder of the paper is organized as follows.
    We briefly review related work in Section~\ref{sec_related}. Section~\ref{sec_preliminaries} presents necessary preliminaries on the problem background. After this, we detail the framework for the proposed model in Section~\ref{sec-method}. Section~\ref{sec_experiment} demonstrates the experiments conducted to evaluate the utility of the proposed method with discussions. 
    Finally, we conclude this work and present some directions for future work in Section~\ref{sec-conclusion}.
    \end{spacing}
	
    \vspace{-4mm}
    \section{Related Work}\label{sec_related}
    \begin{spacing}{2.0}
    
    \vspace{-2mm}	
    \subsection{Popularity Bias and Debias Approach}
    Popularity bias has been a pervasive problem within contemporary recommender systems, leading to monopolies by a few prominent brands and diminishing consumer satisfaction and engagement. To mitigate the harmful influence stemming from both the data and the model perspectives, a range of debiasing methods have been proposed, which can be roughly classified into three categories based on their processing stage: pre-processing, in-processing, and post-processing schemes. Pre-processing approaches aim to address the data bias by balancing the skewed distribution of the training data \cite{xv2022neutralizing, calmon2017optimized}. For example, Calmon et al. \cite{calmon2017optimized} designed a probabilistic data pre-processing method to alleviate popularity bias.

    In-processing approaches are designed to mitigate the bias that a model may develop from the imbalanced data it is trained on. 
    Inverse propensity scoring (IPS) \cite{steck2011item, schnabel2016recommendations} is a common strategy that assigns weights to each item according to its popularity. 
    Causal inference \cite{wei2021model, zheng2021disentangling} employs causal graphs to capture and remove the causal effect that item popularity has on the predicted recommendation scores. 
    TIDE \cite{zhao2022popularity} tried to disentangle the benign and harmful parts of popularity bias with temporal information and exploits the benign bias to enhance recommendation accuracy.
    MACR \cite{wei2021model} formulate a causal graph to describe the important cause-effect relations in the recommendation process. 
    AdaSIR \cite{chen2022learning} employs an adaptive sampler to select negative samples from a sampling pool, calculating sampling weights based on the model's current state and a fixed popularity distribution.
    Rhee et al. \cite{rhee2022countering} propose to extend the BPR loss with a regularization term that tries to minimize the score differences within positive and negative items, respectively, thus reducing model bias while maintaining high accuracy.
    However, the regularization should be carefully designed and used, or it might exacerbate the popularity bias. 
 
    Post-processing approaches offer a more direct approach compared to the other two schemes, as they directly adjust the biased score/rank generated by off-the-shelf recommendation algorithms. For instance, xQuAD \cite{abdollahpouri2019managing} incorporates personalized bonuses to less popular items, aiming to elevate their rankings. Some research addresses popularity bias from a user-centric perspective \cite{abdollahpouri2021user}. These studies introduce metrics to re-rank recommendation outcomes under the assumption that recommendations for a particular user follow a similar distribution to their interacted items.  MF-PC \cite{zhu2021popularity} supplements original recommendation outcomes with compensated scores derived from item popularity.

    Among various types of popularity bias, we focus mainly on the model bias. Unlike prevailing debias methodologies, our work proposes a novel debias design by explicitly modeling popularity and separating it with other latent item features by orthogonal projection. This approach maintains the expressiveness of our tensor-based latent factor model and overcomes limitations such as undesired accuracy-debias tradeoff and excessive parameter tuning.	

    \vspace{-2mm}
    \subsection{Data Sparsity and Learning with Auxiliary Data}
    There are two main research lines addressing the extreme sparsity of positive user-item interactions for implicit top-$K$ recommendation. As one category of approaches, latent factor (LF) models aim to assign each user a vector of latent user features and each item a vector of latent item features. These compact latent factors enable similarity computations between arbitrary user-item pairs, thereby addressing the data sparsity issue. As a representative example of LF models, matrix factorization (MF) methods model the preference of user $u$ over item $i$ as the inner product of their latent representations \cite{koren2009matrix}. Various works can be seen as a variation of MF methods by employing different models to learn the user-item interaction function \cite{he2018pseudo} or using different loss functions \cite{rendle2012bpr}.

    Besides improving the recommendation model itself, another line of research tries to enrich the scarce positive interactions by incorporating auxiliary information, sometimes referred to as side information. The side information may come in various forms and from different sources, such as social network relationships \cite{zhao2014leveraging}, text reviews on items \cite{duan2022combining, guo2020consumer}, diverse user preferences \cite{li2021personalized, ren2024consumer}, and so on. The methods for obtaining side information in practical management applications are different: some algorithms in the literature utilize explicit collection methods \cite{he2018pseudo, bertsimas2023tensor}, while others extract it directly from raw data \cite{farias2019learning, duan2022combining}. For e-commerce businesses, users may generate a large amount of auxiliary feedback (such as click history and collect status) other than the primary (target) purchase feedback. In recent years, utilizing such multi-behavior data has attracted a lot of attention for building a fine-grained recommendation system \cite{ding2020improving, liu2017personalized, chen2020efficient, meng2023hierarchical}. In this paper, we combined these two approaches by integrating the multi-behavior side information into a tensor-based latent factor model, which has not been considered in present debiasing recommendation methods and is also well-suited for our intrinsic popularity debias design.

    \end{spacing}
 
\vspace{-4mm}
\section{Preliminaries}\label{sec_preliminaries}
\begin{spacing}{2.0}
\vspace{-2mm}
In this section, we first detail the problem of top-$K$ recommendation with implicit feedback derived from a single behavioral interaction. Following that, we introduce the broadly adopted matrix factorization scheme for estimating user preference, which serves as an essential component of various top-$K$ recommendation algorithms.

\vspace{-2mm}
\subsection{Top-$K$ Recommendation with Implicit Feedback}
Suppose $U=\left\{1, 2, \ldots, m_1\right\}$ and $V=\left\{1, 2, \ldots, m_2\right\}$ denote the sets of users and items involved in a recommender system, where $m_1=|U|$ and $m_2=|V|$ are the numbers of users and items respectively. Implicit feedback contains the records of user-item interactions of some particular behavioral type, say, whether a customer has purchased a product. In most cases, the collected implicit feedback datasets only consist of binary interaction logs. Therefore, we focus on binary interactions in this paper, while non-binary interactions (such as multiple purchase records on a single product from one customer over some time period) can be handled with slight modifications to suit our settings. Let $X=\left\{x_{u v} \mid u \in U, v \in V\right\}\in\{0,1\}^{m_1 \times m_2}$ denote the observed user-item interaction matrix, where $x_{u v}=1$ implies that user $u$ has interacted with item $v$, and $x_{u v}=0$ otherwise. Based on this binary observation matrix $X$, the implicit top-$K$ recommendation aims to create a personalized list of $k$ items for every user $u$, where each item is ranked by the (hidden) preferences of $u$. Hence, the essence of algorithmic design lies in the accurate estimation of the user preference matrix, denoted as $\hat{X}=\left\{\hat{x}_{u v} \mid u \in U, v \in V\right\} \in \mathbb{R}^{m1 \times m2}$. The preference matrix entry $\hat{x}_{u v}$ represents the propensity of user $u$ for item $v$, crucial for personalized recommendations.

\vspace{-2mm}
\subsection{Matrix Factorization for Preference Estimation}\label{sec_MFPE}
\begin{figure}[t]\vspace{2mm}
    \centering
    \includegraphics[width=0.9\linewidth]{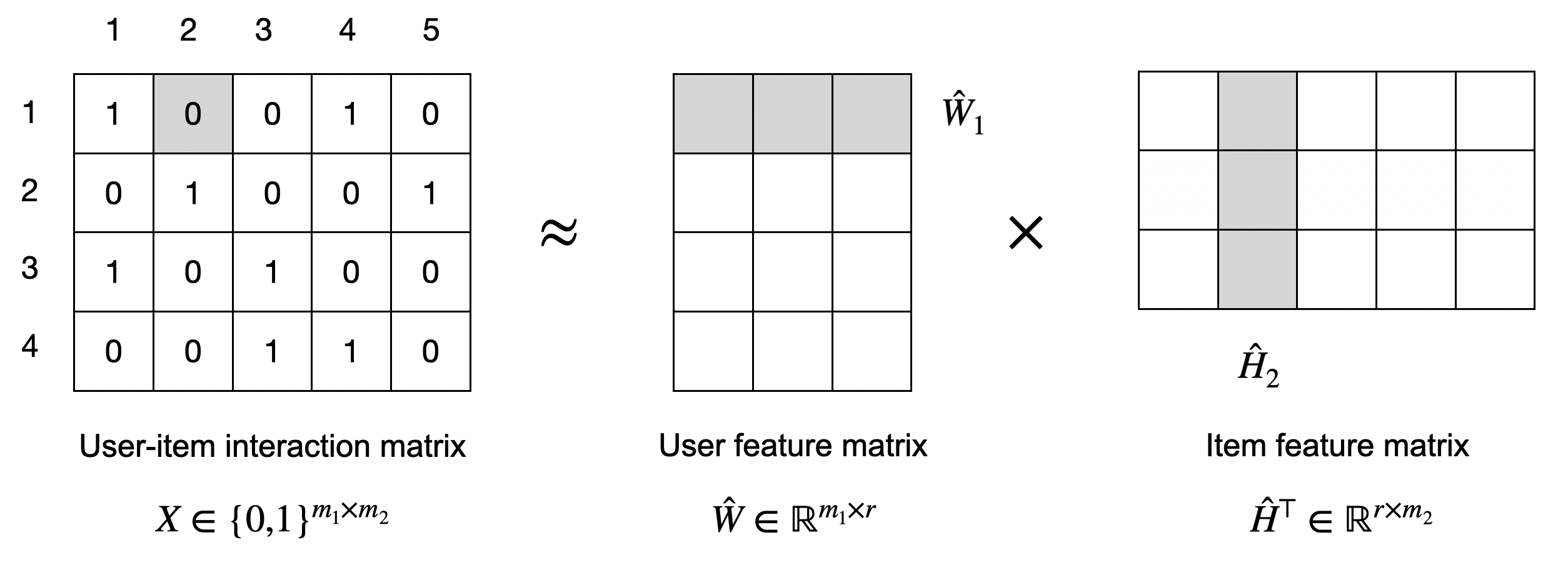}
    \caption{A toy example of matrix factorization scheme on the binary implicit feedback data.}\label{fig-MF}
\end{figure}

A well-established technique for estimating user preference is through latent factor models like matrix factorization \cite{koren2009matrix}, which is still among the most widely used and also the foundation of many state-of-the-art recommendation models \cite{he2018pseudo, he2017neural, rao2015collaborative}. Specifically, suppose the data we have is a binary user-item interaction matrix $X \in\{0,1\}^{m_1 \times m_2}$ encoding historical sales transactions, and the goal for now is to estimate the underlying purchase preference matrix $\hat{X} \in \mathbb{R}^{m_1 \times m_2}$. The matrix factorization formulation builds on the assumption that the latent preference matrix $\hat{X}$ can be decomposed into low-dimensional representations, typically into compact and interpretable features of the users and items. By doing so, the user-item interactions are modeled as inner products in that joint latent feature space. Mathematically, each entry $\hat{x}_{u v}$ of $\hat{X}$ can be estimated as:
\begin{equation}
    \hat{x}_{u v}=<\hat{W}_u, \hat{H}_v>=\hat{W}_u \hat{H}_v^T,
\end{equation}
where $\hat{W}\in \mathbb{R}^{m_1 \times r}$ represents the estimated user latent feature matrix and has its $u$th row $\hat{W}_u$ encoding latent feature of one particular user $u$. Similarly, $\hat{H}\in \mathbb{R}^{m_2 \times r}$ and $ \hat{H}_v$ represent the estimated item feature matrix and latent feature for item $v$. The parameter $r$ is the dimension of joint low-rank latent feature spaces, playing a critical role in the algorithmic performance and usually requiring careful tuning on different tasks. A toy example of such matrix factorization is illustrated in Figure~\ref{fig-MF}. To formulate the matrix factorization model and learn the model parameters, one typically looks to solve the following regularized, squared reconstruction error optimization program \cite{hu2008collaborative, he2016fast}:
\begin{equation}\label{eq-MF}  
    \hat{W}, \hat{H}=\arg \min _{W, H} \frac{1}{2}\left\|S \odot (X-W H^\top\right)\|_F^2+\frac{\lambda_W}{2}\|W\|_F^2+\frac{\lambda_H}{2}\|H\|_F^2,
\end{equation} 
where $\lambda_W$ and $\lambda_H$ are the balancing parameters of corresponding regularizers, which are used to prevent over-fitting. $S \in \mathbb{R}^{m1 \times m2}$ is a weight matrix encoding the weight of each entry and $\odot$ is the Hadamard product operator (i.e. element-wise product) between matrices of conforming dimensions. 
In recommender systems where explicit ratings are given, the weight matrix works like a projection operator $P_{\Omega}$, where $\Omega$ denotes the indices of the observed entries. This projection operator $P_{\Omega}$ projects missing entries as zero while observed entries as themselves, meaning that Eq.\eqref{eq-MF} models directly only on the observed ratings \cite{rao2015collaborative}. 
However, in cases where implicit feedback is involved, the negative signal for user preferences over items is naturally scarce, and it is hard to distinguish between positive (liked) and negative (disliked) instances for each user. Hence, missing entries are usually assigned a zero value but a non-zero weight, corresponding to different negative sampling strategies \cite{chen2023revisiting}.

\end{spacing}

\vspace{-4mm}  
\section{Popularity-Aware Recommendation Framework}\label{sec-method}
\begin{figure}[t]\vspace{0mm}
		\centering
		\includegraphics[width=\linewidth]{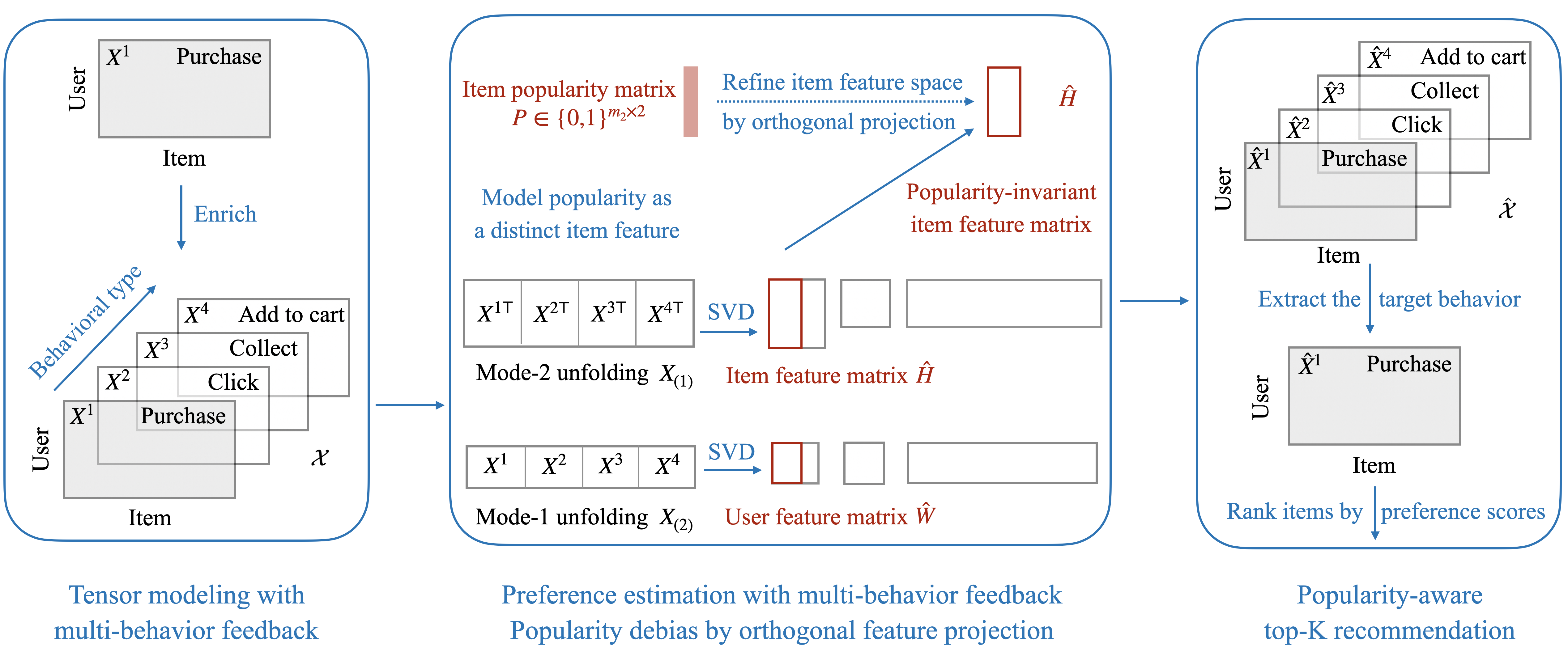}
		\caption{The illustration of the proposed popularity-aware recommendation algorithm PopSI.}\label{framework}
	\end{figure}

\begin{spacing}{2.0}
\vspace{-2mm}
In this section, we present our \textbf{Pop}ularity-aware top-$K$ recommendation algorithm integrated with multi-behavior \textbf{S}ide \textbf{I}nformation (PopSI). A reliable top-$K$ recommendation method requires both accurate estimation of user preferences and the ability to mitigate the influence of popularity bias with a minimal tradeoff in recommendation accuracy. The main rationale guiding our algorithmic design stems from the belief that leveraging multi-behavior feedback can lead to a more precise estimation of the inherent latent features, while the creation of a popularity-invariant item feature space could contribute notably to the debias performance. Illustrating with the motivating example of product recommendation in e-commerce businesses, Figure \ref{framework} provides an overview of the proposed popularity-aware recommendation framework, which contains two crucial parts:
	\begin{itemize}
		\item To address the extreme sparsity exhibited in the observed user-item purchase matrix, we first collect various types of user-item interactions to enrich the original data. By representing these interaction matrices as slices of a three-dimensional tensor $\mathcal{X}$ and assuming the latent user preference tensor $\hat{\mathcal{X}}$ exhibits low slice rank, we can utilize all slices of the multi-behavior data $\mathcal{X}$ to help estimate the desiring purchase slice, whose entries capture crucial signals of customers' hidden purchase tendencies.
  
		\item Following that, we introduce a novel debias design into the established latent factor model. First, we integrate popularity as an explicit and separate item feature through one-hot encoding. Then, we refine the estimated item feature space by making it orthogonal to the item popularity feature vectors. This deliberate orthogonality constraint on item feature space addresses the sensitivity of our method to popularity bias while maintaining the expressiveness of the other latent factors, thereby ensuring a more reliable and unbiased recommendation result.
	\end{itemize}
  
\vspace{-2mm}
\subsection{Preference Estimation with Multiple Behavior} \label{preference}
As stated in Section \ref{sec_MFPE}, matrix factorization stands as a prevalent model for estimating user preferences within modern recommender systems. Nevertheless, when dealing with extremely sparse user-item interactions, like those found in historical sales transaction records from e-commerce businesses, merely applying such methodology oftentimes yields inaccurate and biased outcomes. This is due to the scarcity of positive user-item pairs ($x_{u i}=1$), which hinders the model from deriving meaningful latent factors \cite{he2018pseudo}.
Fortunately, beyond the target sales transactions, online e-commerce platforms have access to a variety of diverse user-item interactions, such as click history and add-to-cart status. These auxiliary interactions typically process greater density compared to the primary purchase records. Though indicating different behavioral preferences, these multi-behavior data are highly correlated and have the potential to reveal similar inherent characteristics of the users and items \cite{farias2019learning}. 
Consequently, leveraging these multi-behavior data allows for a more comprehensive characterization of the target purchase preference and ultimately helps resolve the challenge of limited data. 

Before delving into our model formulation and further discussion, we first briefly present some necessary tensor basics. For an overview of tensor decompositions and applications, we refer the readers to the comprehensive review \cite{kolda2009tensor}. Tensors can be viewed as high-dimensional extensions of matrices and vectors. Consider a three-dimensional tensor $\mathcal{X} \in \mathbb{R}^{m_1 \times m_2 \times n}$, it can be seen as a series of $n$ matrices, denoted as $X^1, \ldots, X^n \in \mathbb{R}^{m_1 \times m_2}$, stacking along the third dimension. These matrices are called the frontal slices of $\mathcal{X}$, which in our context, represent various user-item interactions, each for one particular behavior type. 
One frequently used tensor operation throughout this paper is \textit{unfolding}, which is the process of reordering the elements of the tensor $\mathcal{X}$ into a matrix. The mode-$k$ unfolding of a tensor $\mathcal{X}$, denoted by $X_{(k)}$, arranges the mode-$k$ fibers to be the columns of the resulting matrix. Tensor element $\left(i_1, i_2, i_3\right)$ are mapped into matrix element $\left(i_k, j\right)$, where
\begin{equation}
    j=1+\sum_{\substack{p=1 \\ p \neq k}}^3\left(i_p-1\right) J_p \quad \text { with } \quad J_p=\prod_{\substack{q=1 \\ q \neq k}}^{p-1} I_q.
\end{equation}
Specifically, the columns of the mode-1 unfolding $X_{(1)}\in\mathbb{R}^{m_1 \times m_2 n}$ of tensor $\mathcal{X} \in \mathbb{R}^{m_1 \times m_2 \times n}$ are equivalent to the columns of all of the slices in the original tensor. Similarly, the columns of the mode-2 unfolding $X_{(1)}\in\mathbb{R}^{m_2 \times m_1 n}$ are equivalent to the rows of all of the slices in the original tensor.
Visualizations of a three-dimensional tensor $\mathcal{X}$ and its model-1 and mode-2 unfoldings $X_{(1)}$ and $X_{(2)}$ can be seen in Figure~\ref{framework}.
Another thing required is the concept of tensor \textit{rank}, which is analogous to and the generalization of matrix rank. However, the definition of tensor rank is not unique \cite{kolda2009tensor}. 
In this paper, we adopt the newly proposed slice rank \cite{farias2019learning} for three-dimensional tensors, whose benefits will be clarified in the sequel. The slice rank of a three-dimensional tensor $\mathcal{X}$, denoted as Slice$(\mathcal{X})$, is the maximum of the ranks of its mode-1 and mode-2 unfoldings, i.e. Slice $(\mathcal{X})=\max \left(\operatorname{rank}\left(X_{(1)}\right), \operatorname{rank}\left(X_{(2)}\right)\right)$.

Suppose we can observe multiple types of user-item interactions beyond the purchase matrix $X \in \{0,1\}^{m_1 \times m_2}$, and together we represent it as a three-dimensional tensor $\mathcal{X} \in \{0,1\}^{m_1 \times m_2 \times n}$ with each frontal slice representing a specific user-item interaction, encompassing user behaviors such as purchase, click, collection, and add-to-cart, as illustrated in Figure \ref{framework}. 
Clearly, it is infeasible to recover the underlying feature tensor $\hat{\mathcal{X}}$ from $\mathcal{X}$ without any structural assumptions on the observed data. Similar to the low-rank assumption in the matrix case, we assume that there are only a handful of latent user and item features across all interactions. 
To exploit the inherent inter-connected structure within these multiple user behaviors, we assume a specific bi-linear latent factor model for each behavioral type of interaction, which can be formulated as
\begin{equation}\label{eq-TF}
   X^k \approx W S^k H^{\top}, \quad k = 1, 2, \dots, n.
\end{equation}
Here, $S^k$ is an $r \times r$ matrix capturing specific bi-linear user-item relationship for each interaction record $X^k$. $W$ and  $H$ are $m_1 \times r$ and  $m_2 \times r$ matrices, whose rows represent user and item latent features respectively. Notice here that the latent features are the same across slices, which relates various interactions $X^k$ to each other and allows for the potential of using other slices for learning one particular slice, which in our case, is the purchase matrix $X^1$. On the other hand, $S^k$ may vary on different slices, as diverse interactions could involve different features or with different weights, which gives this model potential generality, controlled by $r$, the dimension of the latent feature space. This factorization scheme \eqref{eq-TF} extends the one proposed in \cite{nickel2011three} for three-dimensional tensors with symmetric slices and corresponds to assuming the underlying preference tensor $\hat{\mathcal{X}} \in \mathbb{R}^{m_1 \times m_2 \times n}$ with a slice rank of $r$ \cite{farias2019learning}.

To find $W$, $S^k$, and $H$, we consider to solve the following non-convex, squared reconstruction error optimization program:
    \begin{equation}\label{eq_TFOP}  
         \hat{W}, \hat{S}^k, \hat{H} = \arg \min_{W, S^k, H}  \sum_{k=1}^{n} \|X^k - W S^k H^\top\|_F^2.
    \end{equation} 
Considering the non-convexity nature of the above optimization problem \ref{eq_TFOP}, we cannot get the global optimal solution easily. However, we can still find good approximations via non-convex optimization techniques, and one common strategy tackling it is to use alternating least-squares (ALS) as in \cite{nickel2011three}.
In this study, we adopt an iterative procedure based on the recently proposed slice learning (SL) algorithm \cite{farias2019learning}. The SL algorithm is both computationally efficient that scales to enormous data sizes and theoretically achieves the minimax lower bound for preference estimation with sufficiently many slices. Furthermore, the procedure is particularly well-suited for our downstream task of popularity debias design. The SL algorithm proceeds in two stages, estimating feature subspaces $W$ and $H$, and then learning bi-linear relationship $S^k$ for each slice, together leading to the estimation of user preferences.

\textbf{Estimating Feature Subspaces:} Firstly, we generate estimations for the feature matrices $W$ and $H$ by utilizing all tensor slices.
The procedure for learning $W$ and $H$ shares similarities, therefore, we elaborate on the specifics of estimating item feature matrix $H$ here. Noticing that each column of $H$ encodes one particular latent item feature, estimating $H$ involves approximating its column space. Under mild conditions, we can expect that the columns of $X_{(2)}$ lie approximately in $H$ \cite{farias2019learning}. Subsequently, we can derive $\hat{H}$ as the first $r$ left singular vectors of $X_{(2)}$ as a good approximation, which is denoted as $\operatorname{svds}\left(X_{(2)}, r\right)$. Similarly, $\hat{W}$ can be estimated as the first $r$ left singular vectors of $\mathcal{X}_{(1)}$ as $\operatorname{svds}\left(X_{(1)}, r\right)$.

\vspace{-2mm}
\subsection{Popularity Debias through Orthogonality Constraint}
The engagement of a user with an item is not solely determined by her personal preference, it is also influenced by the inherent features of the item. Popularity, being among the most intuitive and prominent item features, has a notable influence on real-world recommender systems. In practice, users' actual preferences are malleable and can be significantly influenced by the popularity of items, owing to the conformity phenomenon \cite{canamares2018should} and the anchoring effect \cite{adomavicius2013recommender}. Therefore, in addition to ensuring an accurate estimation of the underlying feature spaces, as elaborated in Section \ref{preference}, a reliable recommendation algorithm must also demonstrate sensitivity toward item popularity. 
 
To mitigate the influence of popularity bias arising from both the imbalanced distribution of item popularity and the inherent model bias, prevailing debias methodologies commonly employ an implicit strategy. This involves adjusting biased preference scores inversely proportional to item popularity, a practice that frequently leads to a notable tradeoff between recommendation accuracy and debias performance. 
As demonstrated theoretically in \cite{zhu2021popularity}, the matrix factorization model can produce popularity bias. This originates from the fact that the estimated item feature matrix implicitly encodes item popularity, which inadvertently introduces a bias towards more popular items. To address this issue, we propose a debias design that explicitly represents popularity as a separate and distinct item feature. Subsequently, by disentangling it from other implicit item features via orthogonal feature projection, we aim to mitigate the popularity bias within our multi-behavior integrated latent factor model while maintaining the accuracy of recommendation.
	
\vspace{-2mm}
\subsubsection{Representing Popularity as an Explicit and Distinct Item Feature}
The popularity of an item is conventionally quantified as the aggregate number of interactions it gathers across all users. In our motivating example of product recommendation, this metric is specifically represented as the number of purchases, denoted as $pop (\cdot)$. 
We designate the top $p$ percentage of items based on their purchase frequency as ``popular", while the remaining $1-p$ percentage are classified as ``less popular". Typically, we can artificially set $p$ as 20\% and this division aligns with the established practices within recommender systems \cite{abdollahpouri2017controlling}, striking a balance by emphasizing extensively preferred items while acknowledging the variety among less engaged ones. 
To formally express this concept in mathematical terms, we can generate a two-column matrix $P \in \{0,1\}^{m_2 \times 2}$ to explicitly represent the item popularity feature through one-hot encoding. This matrix distinctly signifies whether an item falls within the `popular' or `less popular' category, enabling the recommender system to consider and accommodate varying levels of item popularity in its suggestions.

 \vspace{-2mm}
\subsubsection{Disentangling Popularity via Orthogonal Feature Projection}
As we have clarified in Section \ref{preference}, each column in the estimated item feature matrix $\hat{H}\in \mathbb{R}^{m_2 \times r}$ can be seen as a specific, potentially meaningful implicit item feature. Despite the model's clarity in interpretation, one significant obstacle in debias design emerges from the potential distribution of popularity-related information across different dimensions within the item feature space, which in the end affects the estimated preference scores implicitly.
	To integrate the popularity feature matrix $P$ into our model and mitigate harmful bias stemming from item popularity, we propose an additional orthogonality constraint to refine the item feature matrix $\hat{H}$ to be irrelevant to $P$. 
  Specifically, the estimated item feature space $\hat{H}$ should exhibit orthogonality to the sensitive subspace spanned by the columns comprising $P$. This can be done in many ways, the debias approach we choose for achieving this involves a process where the recently estimated $\hat{H}$ matrix is projected onto the orthogonal complement of the column space defined by $P$, which is accomplished as 
\begin{equation}
    P_{\mathrm{ran}(P)}(\hat{H}) = \left(I-P\left(P^T P\right)^{-1} P^T\right)\hat{H}=\hat{H}-P\left(P^{\top} P\right)^{-1} P^{\top} \hat{H}.
\end{equation}
   This ensures that the learned item features align themselves in a space that is independent and uncorrelated with the popularity factors encapsulated within $P$. This novel orthogonality constraint on item feature space addresses the sensitivity of our method to popularity bias while maintaining the expressiveness of the other latent factors, thereby ensuring a more reliable and unbiased recommendation result. 

	\textbf{Learning Bi-linear Relationships:} After obtaining the refined popularity-invariant item feature space $\hat{H}$, we proceed to the second stage of SL algorithm for estimating $S^k$, which operates on each slice independently, entailing the resolution of the following program:
	\begin{equation}
	    \hat{S}^k=\underset{S}{\operatorname{argmin}}\left\|\hat{W} S \hat{H}^{\top}-X^k\right\|_F, \quad k = 1, 2, \dots, n.
	\end{equation}
	The above optimization is a standard least-squares problem and thus admits a closed-form solution. Assuming that the columns of $\hat{W}$ and $\hat{H}$ are orthonormal, the solution can be computed as $\hat{S}^k=\hat{W}^{\top} X^k \hat{H}$. Consequently, our final estimate of user preferences $\hat{X}^k$ is
	\begin{equation}
	\hat{X}^k=\hat{W} \hat{W}^{\top} X^k \hat{H} \hat{H}^{\top}, \quad k = 1, 2, \dots, n.
	\end{equation}
 To summarize the above, the scheme for our popularity-aware top-$K$ recommendation algorithm is outlined in Algorithm \ref{algorithm_outline}, where personalized lists are generated based on the preference scores in the estimated purchase preference $\hat{X}^1$.

 	\begin{algorithm}[t]
		\caption{\textbf{Pop}ularity-aware top-$K$ recommendation with \textbf{S}ide \textbf{I}nformation (PopSI)}
		\label{algorithm_outline}
		\begin{algorithmic}[1]
    \begin{spacing}{2.0}
    \vspace{1em}
			\REQUIRE $\mathcal{X}$: observed multi-behavior data; $r$: slice rank;  $ P$: popularity feature matrix.
			\ENSURE $\hat{X}^k, k=1, \ldots, n$: the estimated debiased user-item preference matrices.
			\STATE Compute the matrix SVD of the mode-1 unfolding matrix $X_{(1)}$ to obtain the user feature matrix as the ﬁrst r left singular vectors: $\hat{W} \leftarrow \operatorname{svds}\left(X_{(1)}, r\right)$;
			\STATE Compute the matrix SVD of the mode-2 unfolding matrix $X_{(2)}$ to obtain the item feature matrix as the ﬁrst r left singular vectors: $\hat{H} \leftarrow \operatorname{svds}\left(X_{(2)}, r\right)$;
			\STATE Refine the item feature space by orthogonal projection: $\hat{H} \leftarrow \hat{H}-P\left(P^{\top} P\right)^{-1} P^{\top} \hat{H}$;
			\STATE Get orthonormal basis of $\operatorname{range}(\hat{H})$: $\hat{H} \leftarrow \operatorname{qr}\left(\hat{H}\right)$;
			\STATE Compute the estimated debiased preference matrices: $\hat{X}^k=\hat{W} \hat{W}^{\top} X^k \hat{H} \hat{H}^{\top}, k=1, \ldots, n$.
   
              \noindent\textbf{Return:} $\hat{X}^k=\hat{W} \hat{W}^{\top} X^k \hat{H} \hat{H}^{\top}, k=1, \ldots, n$.
    \end{spacing}
    \vspace{-4mm}
		\end{algorithmic}
	\end{algorithm}

	\end{spacing}

\vspace{-4mm}    
\section{Experiment and Discussion}\label{sec_experiment}
\begin{spacing}{2.0}
\vspace{-2mm}
We conducted a series of experiments to evaluate the popularity-aware top-$K$ recommendation algorithm PopSI on real-world e-commerce datasets. This section describes these experiments and discusses their results in detail, wherein the main findings are as follows:
\begin{enumerate}
\item The PopSI algorithm consistently improves the recommendation accuracy and debias performance compared to other baselines and debias algorithms on various datasets. The improved recommendation accuracy demonstrates the value of utilizing multi-behavior side information for capturing authentic user preferences with an additional contribution to the debias performance by the ablation study. The novel orthogonality constraint we added on the item feature space prevents our model from the influence of popularity bias while introducing a marginal accuracy-debias tradeoff, which is an uncommon occurrence in other debias methods.
\item Unlike conventional debias approaches, which require additional model assumptions or involve extensive hyper-parameters tuning, PopSI has only two parameters that need to be adjusted. The first is the slice rank $r$ that controls the model's overall performance, while the second is the percentage $p$ in the popularity feature matrix $P$ that demonstrates the model's sensitivity to item popularity. This makes our algorithm easy to tune and adaptable to various types of datasets.
\item Conducting primarily singular value decomposition on sparse matrices and orthogonal projections, PopSI requires drastically fewer computations without additional training, which makes our algorithm scalable and applicable to large real-world applications.
\end{enumerate}
    
\subsection{Experimental Setup}
 \subsubsection{Datasets Description}
\textbf{Tmall}. The Tmall dataset\footnote{\url{https://tianchi.aliyun.com/dataset/46}} contains anonymized mobile shopping logs of customers on Tmall.com. Originally, the dataset contains four different types of consumer-product interactions with implicit feedback, namely click, collect, add-to-cart, and purchase. In real-world business scenarios, there is often a need to develop a personalized recommendation system focused on a specific subset of all products. Thus, we generate two subsets containing all four behavior types with different sizes and purchase sparsity for our model evaluations.

\textbf{Beibei}. Beibei\footnote{\url{http://www.beibei.com/.}} is the largest e-commerce platform for maternal and infant products in China. Originally, the dataset contains three types of user behaviors, including click, add-to-cart, and purchase. As in the Tmall dataset, we select two subsets of all three interaction types containing consumers and products with more than 5 and 10 purchase interactions.

The primary (target) behavior of our top-$K$ recommendation task is the purchase interaction, as it is a strong signal of a consumer's propensity for a product but is often performed with the least frequency among all behavioral types.
It is worth noting that the user behavior is timestamped down to the hour, so it's possible to have completely identical behaviors within an hour. In our study, we transform all types of interactions into binary entries. The detailed statistical summary of the chosen datasets is presented in Table \ref{subset details}, where we displayed the number of interactions within the corresponding behavioral type in each dataset, along with the purchase sparsity. As can be observed, these subdatasets exhibit a high level of sparsity on the purchase interaction slice, even as low as $0.03\%$.

\begin{table}[t]
\centering
\begin{spacing}{1.5}
\caption{Statistical details of the evaluation datasets.}\label{subset details}
\end{spacing}
\begin{tabular}{cccccccc}
\toprule
Dataset  & \# Users & \# Items & \# Click & \# Collect & \# Cart & \# Purchase & Sparsity \\\midrule
Tmall-S  & 1463     & 800      & 8938     & 456        & 1533    & 1665        & 0.14\%   \\
Tmall-L  & 8562     & 6412     & 118905   & 6499       & 15110   & 16588       & 0.03\%   \\
Beibei-S & 15660    & 4544     & 1426647  & -          & 421650  & 238616      & 0.34\%   \\
Beibei-L & 21716    & 7977     & 2412586  & -          & 642622  & 304576      & 0.18\%   \\
\bottomrule
\end{tabular}
\end{table}
	
For our experiments, we partitioned each dataset randomly into three parts: $80 \%$ for training, $10 \%$ for validation (for parameter tuning only), and $10 \%$ for testing (for evaluation). Given that many consumers exhibit minimal interactions, it's possible that they may not appear in the testing set. Consequently, there could be no alignment between the top-$K$ recommendation lists generated by any algorithm and the ground truth, owing to the absence of further purchases by these consumers. The results presented in the sequel contain these instances, reflecting the inherent challenges of recommender systems operating within highly sparse real-world environments.
\vspace{-2mm}
\subsubsection{Baselines and Parameter Setting}
To illustrate the utility of the proposed popularity-aware top-$K$ recommendation method PopSI, we compare it with the following baseline models and state-of-the-art debias methods, which can be classified into three categories. A summary can be seen in Table \ref{algo_type}.

The compared single-behavior methods without debias design:
\vspace{-2mm}
	\begin{itemize}
		\item \textbf{Itempop}. This simple method recommends the most popular products equally to all consumers, where all products are ranked by the number of times they have been purchased within our context.
		\item \textbf{MF} \cite{koren2009matrix}. This is a representative collaborative filtering model using the matrix factorization (MF) technique, which uses mean square error as the objective function with negative sampling from non-interacted entries.
		\item \textbf{BPR} \cite{rendle2012bpr}. Bayesian personalized ranking (BPR) is a standard ranking framework for implicit recommendation that uses pairwise Bayesian personalized ranking loss and samples negative items with uniform distributions.
  \end{itemize}
  
  The compared multi-behavior methods without debias design:
  \begin{itemize}
            \item \textbf{CML} \cite{wei2022contrastive}. This method devised a contrastive meta learning (CML) model for multi-behavior recommendation by preserving behavior heterogeneous context with the agreement between behaviors views.
            \item \textbf{MBGMN} \cite{xia2021graph}. This is a multi-behavior recommendation framework that utilizes graph meta network to incorporate the multi-behavior into a meta-learning paradigm.
            \end{itemize}
            
              The compared single-behavior methods with debias design:
            \begin{itemize}
		\item \textbf{MFW} \cite{steck2011item}. This method removes the popularity bias by weighted matrix factorization. The weights assigned to training samples in the recommendation loss depend on the popularity of the items involved. Items with low popularity are assigned high weights, which helps to increase the predicted scores for them.
		\item \textbf{MFR} \cite{steck2019collaborative}. This method removes popularity bias by rescaling the training data, which multiplies rescaling values to the binary training samples based on the popularity of the involved items to uniformly promote the scores of low-popularity items.
		\item \textbf{MF-PC} \cite{zhu2021popularity}. This is a state-of-the-art popularity compensation (PC) rank adjustment method for mitigating popularity bias. It has two choices: re-ranking and regularization. Here we implement the re-ranking method based on the BPR loss because this method shows better performance in the original paper. We tune the bias-controlling hyper-parameters $\alpha\in[0.1,1.5]$ with step 0.1, and $\beta\in[0,1]$ with step 0.1.
		\item \textbf{Zerosum} \cite{rhee2022countering}. Zerosum is a regularization-based method that extends the widely used BPR loss with a regularization term that minimizes the score differences within preferred and unpreferred items, respectively. The weight of the term is fixed at 0.1 for all experiments.
	\end{itemize}

\begin{table}[t]
\centering
\begin{spacing}{1.5}
\caption{A summary of competing methods based on their utilization of multi-behavior and debias design.}\label{algo_type}
\end{spacing}
\begin{tabularx}{0.8\linewidth}{XXXX}
\hline
Method        &Single-behavior & Multi-behavior & Debias design  \\ \hline 
ItemPop       &$\checkmark$    &$\times$        &$\times$        \\
MF            &$\checkmark$    &$\times$        &$\times$        \\
BPR           &$\checkmark$    &$\times$        &$\times$        \\
\hline
CML           &$\checkmark$    &$\checkmark$    &$\times$        \\
MBGMN         &$\checkmark$    &$\checkmark$    &$\times$        \\
\hline
MFW           &$\checkmark$     &$\times$       &$\checkmark$    \\
MFR           &$\checkmark$     &$\times$       &$\checkmark$    \\
MF-PC         &$\checkmark$     &$\times$       &$\checkmark$    \\
Zerosum       &$\checkmark$     &$\times$       &$\checkmark$    \\
\hline
PopSI (ours)  &$\checkmark$     &$\checkmark$   &$\checkmark$    \\
\hline
\end{tabularx}
\end{table}
	
In the experiments, we fix the dimension $d$ of users’ and items’ embedding as 128, and the learning rate as 0.001, the negative sampling rate is set to 2. Then we tune hyper-parameters for all models by grid search over validation sets.
Note that for all of the baselines and debias approaches, there is a tradeoff between recommendation accuracy and debias performance. Therefore, we aim to optimize hyper-parameters that minimize the bias metric while preserving an acceptable recommendation accuracy.  
			\vspace{-2mm}
	\subsubsection{Evaluation Metrics}
	The utility of top-$K$ recommendation methods is evaluated from two perspectives: recommendation accuracy and debias performance. To evaluate the recommendation accuracy, we choose two widely used metrics in top-$K$ recommender systems: $Recall@K$ and $NDCG@K$ to evaluate the performance of the proposed methods. $Recall@K$ measures the ratio of truly predicted items in the test set. It is formulated as follows:
	\begin{equation}
	   Recall @ K=\frac{1}{N} \sum_{u \in U} \frac{\left|R_u @ K \cap T_u\right|}{\left|T_u\right|}, 
	\end{equation}
	where $R_u @ K$ denotes user $u$'s top-$K$ recommendation results, and $T_u$ denotes positive items in the test set for $u$. Discounted Cumulative Gain $DCG@K$ is a metric for measuring ranking quality. It considers the effects of different recommended items in a ranking list on the recommendation performance:
        \begin{equation}
	D C G @ K=\sum_{i=1}^K \frac{2^{r e l_i}-1}{\log _2(i+1)},
	\end{equation}
	where $\mathrm{rel}_i$ denotes whether the $i$-th item in the top-$K$ recommendation results is truly predicted. For user $u$, $\mathrm{rel}_i=1$ when the $i$-th item in $R_u @ K$ is in $T_u$ and $\mathrm{rel}_i=0$ otherwise. $N D C G @ K$ is normalized $D C G @ K$, which is formulated as follows:
         \begin{equation}
	N D C G @ K=\frac{1}{N} \sum_{u \in U} \frac{D C G_u @ K}{I D C G_u @ K},
	\end{equation}
	where $I D C G_u @ K$ denotes the ideal $D C G_u @ K$. It is reached when all truly predicted items are ranked top in $R_u @ K$.
	
	Some prior research suggested several metrics to measure the popularity bias \cite{boratto2021connecting, zhu2021popularity}. In our experiments, we choose a metric computing the popularity-rank correlation for items ($PRI$) \cite{zhu2021popularity}. $PRI$ computes the Spearman rank correlation coefficient $(S R C)$ of item popularity and the average ranking position, conditioned on the positive items, given by:
          \begin{equation}
	P R I=-S R C\left( pop(I), \operatorname{avg} \_r a n k(I)\right).
	\end{equation}
	The average rank $\operatorname{avg}\_rank$ of each item $i$ is determined as follows: for each item $i$, we identify users $u$ who include $i$ in their set of positive items $\operatorname{Pos}_u$ and calculate the rank position quantile of $i$ within $\operatorname{Pos}_u$. Subsequently, the rank position quantiles are averaged across all users $u$ who have $i$ in their $\operatorname{Pos}_u$. It's important to note that a smaller average rank quantile, closer to 0, indicates that the item tends to receive higher scores among the $\operatorname{Pos}_u$ of each user, while a larger average rank quantile, closer to 1, suggests that the item generally receives lower scores compared to other positive items. Therefore, a $PRI$ value approaching 1 implies that the model assigns higher scores to more popular items, whereas a $PRI$ value nearing 0 suggests that the model exhibits less bias, as there is little correlation between the popularity of positive items and the ranking of recommendation scores.

\begin{table}[htbp]
\centering
\begin{spacing}{1.5}
\caption{The recommendation accuracy and debias performance results of competing methods on various datasets. The best results are highlighted in bold and sub-optimal results within debias approaches (i.e. except for the first five baselines without debias design) are underlined.}\label{main_results_table}
\end{spacing}
\begin{tabularx}{\textwidth}{XXXXXXX}
\hline
Dataset	&Method &$R@20$ $\uparrow$ &$R@50$ $\uparrow$ &$N@20$ $\uparrow$ &$N@50$ $\uparrow$ &$PRI$ $\downarrow$ \\ 
\hline
\multirow{10}{*}{Tmall-S} 
   & ItemPop  &0.0631&0.1291&0.0048&0.0070&0.9666    \\
   & MF       &0.1022&0.1597&0.0090&0.0123&0.5422    \\
   & BPR      &0.1231&0.1702&0.0095&0.0138&0.5271    \\
   \cline{2-7}
   & CML      &0.1539&0.1952&0.0305&0.0401&0.3417    \\
   & MBGMN    &0.1302&0.1831&0.0281&0.0372&0.3289    \\
   \cline{2-7}
   & MFW      &0.0722&0.1352&0.0050&0.0094&0.3478    \\
   & MFR      &0.0718&0.1411&0.0064&0.0100&0.3089    \\
   & MF-PC  &\underline{0.0823}&\underline{0.1425}&0.0079&0.0101&0.2928    \\
   & Zerosum  &0.0811&0.1420&\underline{0.0090}&\underline{0.0122}&\underline{0.2820}    \\
   & PopSI  &\textbf{0.2308}&\textbf{0.3387}& \textbf{0.0595}&\textbf{0.0633}&\textbf{0.1970}    \\
\hline
\multirow{10}{*}{Tmall-L} 
   & ItemPop  &0.0148&0.0286&0.0018&0.0026&0.8897    \\
   & MF       &0.0230&0.0477&0.0043&0.0076&0.4581    \\
   & BPR      &0.0298&0.0542&0.0050&0.0091&0.4300    \\
   \cline{2-7}
   & CML      &0.0672&0.1053&0.0100&0.0194&0.2817    \\
   & MBGMN    &0.0418&0.0762&0.0084&0.0141&0.2889    \\
   \cline{2-7}
   & MFW      &0.0182&0.0356&0.0032&0.0051&0.2001    \\
   & MFR      &\underline{0.0185}&\underline{0.0412}&0.0039&\underline{0.0058}&0.1878    \\
   & MF-PC    &0.0169&0.0302&\underline{0.0040}&\underline{0.0058}&0.2230 \\
   & Zerosum  &0.0150&0.0300&0.0039&0.0052&\underline{0.1720}    \\
   & PopSI    &\textbf{0.1034}&\textbf{0.1369}&\textbf{0.0172}&\textbf{0.0282}&\textbf{0.1029}    \\
\hline

\multirow{10}{*}{Beibei-S} 
   & ItemPop  &0.1313&0.2339&0.0456&0.0634&0.9991    \\
   & MF       &0.1835&0.3264&0.0530&0.0811&0.6568    \\
   & BPR      &0.2161&0.3931&0.0574&0.0893&0.6510    \\
   \cline{2-7}
   & CML      &0.2872&0.4297&0.0821&0.0913&0.4120    \\
   & MBGMN    &0.3181&0.4960&0.1022&0.1201&0.4291    \\
   \cline{2-7}
   & MFW      &0.1513&0.2871&0.0478&0.0658&0.4298    \\
   & MFR      &0.1610&0.2941&0.0495&0.0702&0.4022    \\
   & MF-PC    &0.1622&0.3081&0.0497&0.0715&\underline{0.3371}    \\
   & Zerosum  &\underline{0.1793}&\underline{0.3169}&\underline{0.0522}&\underline{0.0742}&0.3552    \\
   & PopSI    &\textbf{0.4446}&\textbf{0.5885}&\textbf{0.1512}&\textbf{0.1745}&\textbf{0.2897}    \\
\hline

\multirow{10}{*}{Beibei-L} 
   & ItemPop  &0.1173&0.2108&0.0394&0.0548&0.9985    \\
   & MF       &0.1712&0.3121&0.0502&0.0792&0.6092    \\
   & BPR      &0.1922&0.3546&0.0573&0.0871&0.6188    \\
   \cline{2-7}
   & CML      &0.2821&0.3590&0.0830&0.0921&0.3522    \\
   & MBGMN    &0.3034&0.4121&0.1082&0.1172&0.3682    \\
   \cline{2-7}
   & MFW      &0.1502&0.2632&0.0440&0.0683&0.3762    \\
   & MFR      &0.1509&0.2863&0.0442&0.0701&0.3582    \\
   & MF-PC    &0.1475&0.2883&0.0437&0.0695&0.3021    \\
   & Zerosum  &\underline{0.1521}&\underline{0.3021}&\underline{0.0481}&\underline{0.0726}&\underline{0.2981}   \\
   & PopSI    &\textbf{0.3947}&\textbf{0.5179}&\textbf{0.1361}&\textbf{0.1550}&\textbf{0.2040}    \\
\hline
\end{tabularx}
\end{table}
 
\vspace{-2mm}
\subsection{Performance Comparison}
Table \ref{main_results_table} provides a summary of the utilities of all methods for product recommendation on various datasets in terms of both recommendation accuracy and debias performance. 
In absolute terms, the PopSI algorithm performs very well. PopSI consistently improves both the recommendation accuracy and debias performance compared to other baselines and debias algorithms by a large margin.
The rationale behind this enhancement is that it is difficult for traditional matrix-based approaches to capture authentic user preferences due to the extreme rareness of positive interactions within the real-world data, while our tensor-based method capitalizes on multi-behavior side information, enabling a more comprehensive understanding of user preferences. 
The improved recommendation accuracy of PopSI stems from the ability to attain superior estimations of user and item feature spaces, which is achieved by leveraging various user-item interactions in a tensor-based latent factor model effectively.
PopSI also achieved the best debais performance compared to all other debiasing methods, which demonstrates the effectiveness of our popularity debias design. By modeling popularity as an explicit item feature and refining the estimated item feature space to be orthogonal to the columns of the item popularity feature matrix, the derived popularity-aware item feature space not only prevents our model from the influence of the popularity bias but still processes the ability to together capture hidden user preferences accurately.

Some debias approaches perform poorly, and they do not reach the debias performance of the baseline MFR model with minimal enhancements in the recommendation accuracy. It can be concluded that although some of the state-of-the-art debiased models outperform other methods in the original paper, they cannot exhibit the same superiority on other datasets due to hyper-parameter selections or additional model assumptions. However, PopSI shows stable enhancement for the baseline models and performs best across all datasets. Besides, by fixing the popularity matrix $P$, PopSI only has one extra hyper-parameter $r$ to be tuned manually, making our model easily adapted to other real-world datasets, while existing methods often require extensive hyper-parameter tuning to find an appropriate balance of accuracy and debias performance or additional model training.

\subsection{Ablation Study}
In order to validate the value of multi-behavior side information and the effectiveness of the introduced orthogonality constraint for mitigating popularity bias, we conduct a series of ablation experiments on the Tmall-S and Beibei-L datasets. 
While PopSI is based on tensor representation, the basic idea and algorithm remain applicable to the single-behavior (e.g., purchase) matrix cases. For the ablation experiments, we introduce three adaptations of the PopSI approach: direct application of preference estimation to purchase matrix without orthogonal constraint, direct application of preference estimation to purchase matrix with orthogonal constraint, and application to multi-interaction tensor data but lacking orthogonal constraint. The utility of these variations is summarized in Table \ref{ablation}.

Combining the results from Table \ref{ablation} and Table \ref{main_results_table}, we can find that leveraging multi-behavior side information can improve the recommendation accuracy by a large margin, and our orthogonality constraint indeed improves the debias performance without sacrificing much accuracy. 
Specifically, the debias performance on the matrix case on Tmall-S achieves a suboptimal result out of the three variations and is comparable with the results of debias baselines in Table \ref{main_results_table}, which alone demonstrates the effectiveness of our orthogonality constraint debias design.
The accuracy-debias tradeoff introduced by our orthogonality constraint is marginal, which demonstrates the expressive capability of the refined latent item feature factors in our model.
By leveraging multi-behavior side information alone, the recommendation accuracy is enhanced by a large margin compared with baseline models and state-of-the-art debias approaches. 
Surprisingly, we notice that this variation method also represents a certain degree of debias effect and achieves a suboptimal debias result on Beibei-L, though not as sharp as PopSI. One possible explanation is that by leveraging diverse types of user interactions, the popularity bias inherent in the purchase slice gets diluted, thereby enabling this variation method to exhibit the ability to mitigate popularity bias. This also demonstrates the value of utilizing multi-behavior side information in our popularity-aware recommendation framework.

\begin{table}[t]
\centering
\begin{spacing}{1.5}
    \caption{Values of the multi-behavior side information (SI) and the effectiveness of the popularity orthogonality constraint (Pop) of the PopSI algorithm on the Tmall-S and Beibei-L datasets. The best results are highlighted in bold and sub-optimal results are underlined.}\label{ablation}
\end{spacing}
\begin{tabularx}{\textwidth}{XXXXXXXX}
\hline
& Pop&SI&$R@20$ $\uparrow$         & $R@50$ $\uparrow$        & $N@20$ $\uparrow$     &$N@50$ $\uparrow$       & $PRI$ $\downarrow$           \\ \hline
\multirow{4}{*}{Tmall-S} 
&$\times$&$\times$    &0.1081&0.1882&0.0079&0.0113&0.4917    \\
&$\checkmark$&$\times$&0.1011&0.1802&0.0072&0.0093&\underline{0.3109}    \\
&$\times$&$\checkmark$&\textbf{0.2462}&\textbf{0.3493}&\textbf{0.0621}&\textbf{0.0694}&0.3213    \\
&$\checkmark$&$\checkmark$&\underline{0.2308}&\underline{0.3387}& \underline{0.0595}&\underline{0.0633}&\textbf{0.1970}    \\
\hline
\multirow{4}{*}{Beibei-L} 
&$\times$&$\times$        &0.1179             &0.2338             &0.0334             &0.0517             &0.6223    \\
&$\checkmark$&$\times$      &0.0909             &0.1948             &0.0241             &0.0401             &0.5038    \\
&$\times$&$\checkmark$      &\textbf{0.4046}    &\textbf{0.5274}    &\textbf{0.1364}    &\textbf{0.1572}    &\underline{0.4585}    \\
&$\checkmark$&$\checkmark$  &\underline{0.3947} &\underline{0.5179} &\underline{0.1361} &\underline{0.1550} &\textbf{0.2040}    \\
\hline
\end{tabularx}
\end{table}

\vspace{-2mm}
\subsection{Parameter Sensitivity}
\begin{figure}[t]
	\centering
        \caption{Effect of hyper-parameter slice rank $r$ on the recommendation accuracy (represented by $NDCG@50$) and debias performance (represented by $PRI$) of PopSI on two chosen datasets.}\label{sensi_r}   
	\subfigure[Effects of slice rank $r$ on Tmall-S]
	{
		\begin{minipage}[b]{0.49\linewidth}
			\includegraphics[width=1\linewidth]{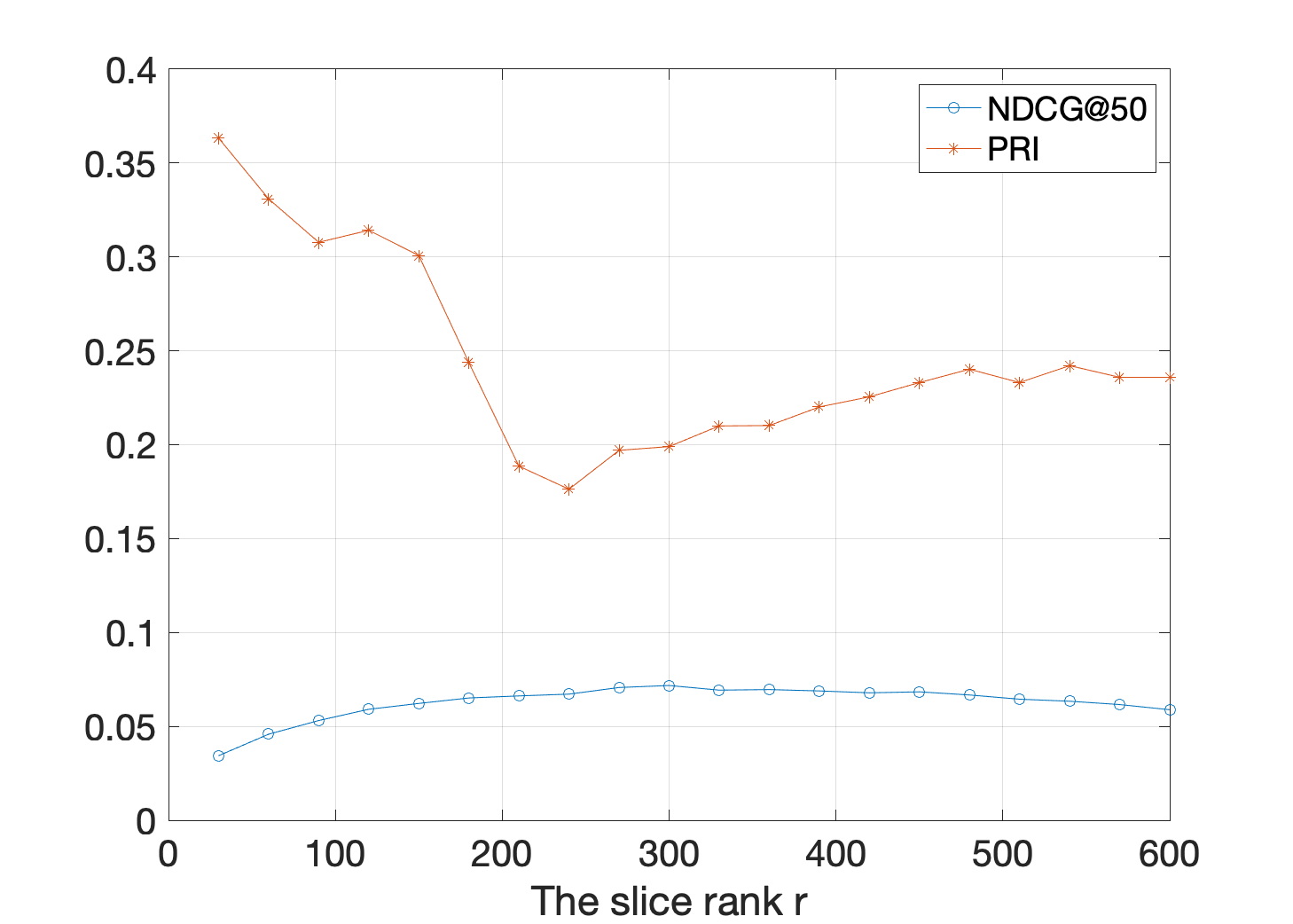}\vspace{-5pt}
		\end{minipage}
	}\hspace{-5mm}
 	\subfigure[Effects of slice rank $r$ on Beibei-L]
	{
		\begin{minipage}[b]{0.49\linewidth}
			\includegraphics[width=1\linewidth]{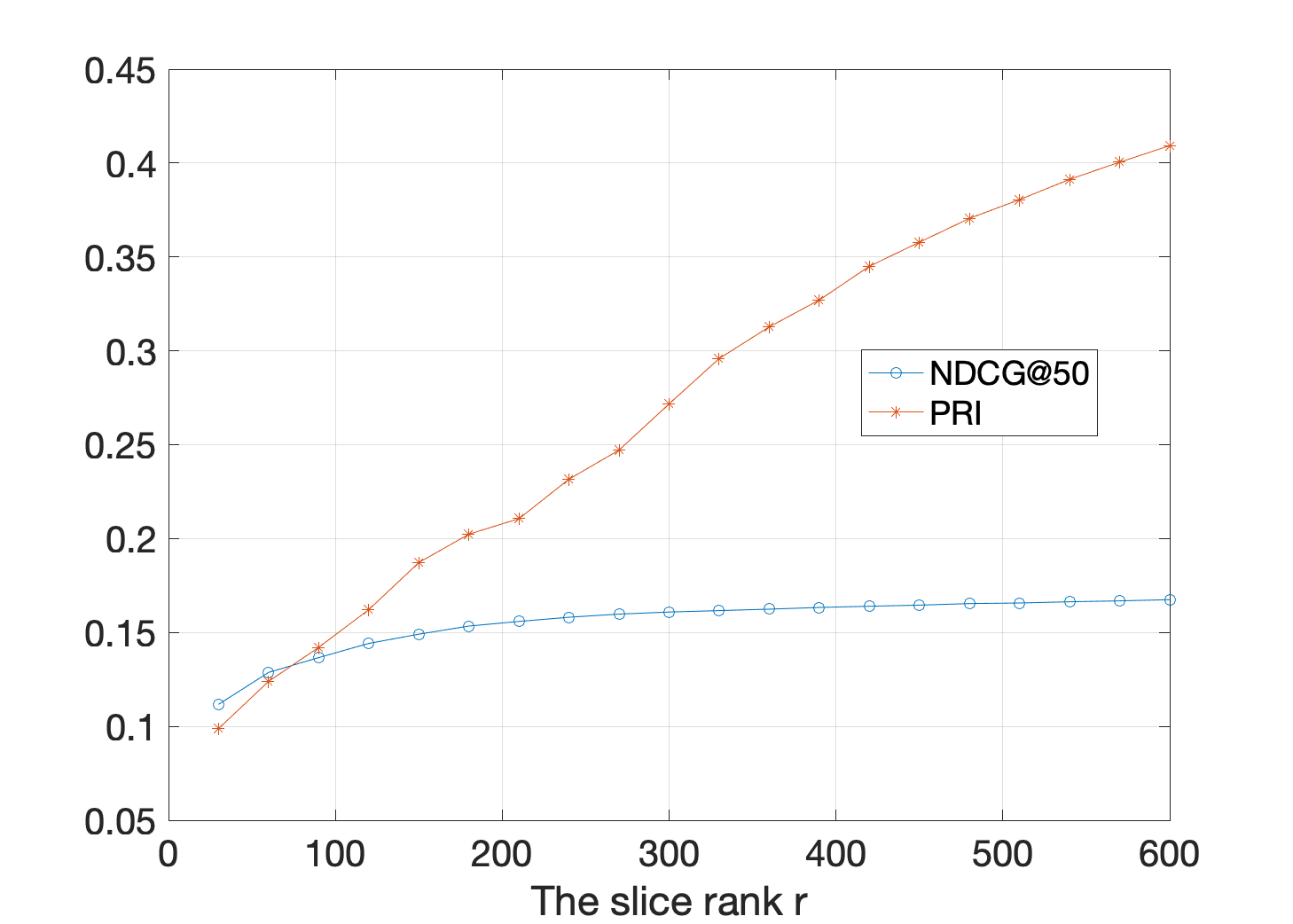}\vspace{-5pt}
		\end{minipage}
	}\hspace{-5mm}
\end{figure}

The proposed popularity-aware recommendation algorithm PopSI has only two hyper-parameters $r$ and $p$, which makes our algorithm easy to tune and adaptable to various types of datasets. 
Note that $r$ is the slice rank of the underlying preference tensor $\hat{\mathcal{X}}$, encoding the dimension of the hidden user and item feature spaces. In Figure \ref{sensi_r}, we illustrate the recommendation accuracy (represented by $NDCG@50$) and debias performance (represented by $PRI$) of PopSI with different values of slice rank $r$ by fixing the percentage $p$ as 0.2 on two chosen datasets, namely Tmall-S and Beibei-L. The larger the value of $r$, the more features the model can represent, potentially enhancing the overall recommendation utility. On the other hand, the value of $r$ can not be overly large, as this might impede the model's ability to simultaneously grasp these latent features accurately, ultimately compromising its accuracy and debias performance. 

\begin{figure}[t]
	\centering
        \caption{Effect of hyper-parameter percentage $p$ on the recommendation accuracy (represented by $NDCG@50$) and debias performance (represented by $PRI$) of PopSI on two chosen datasets.}\label{sensi_p} 
	\subfigure[Effects of percentage $p$ on Tmall-S]
	{
		\begin{minipage}[b]{0.49\linewidth}
			\includegraphics[width=1\linewidth]{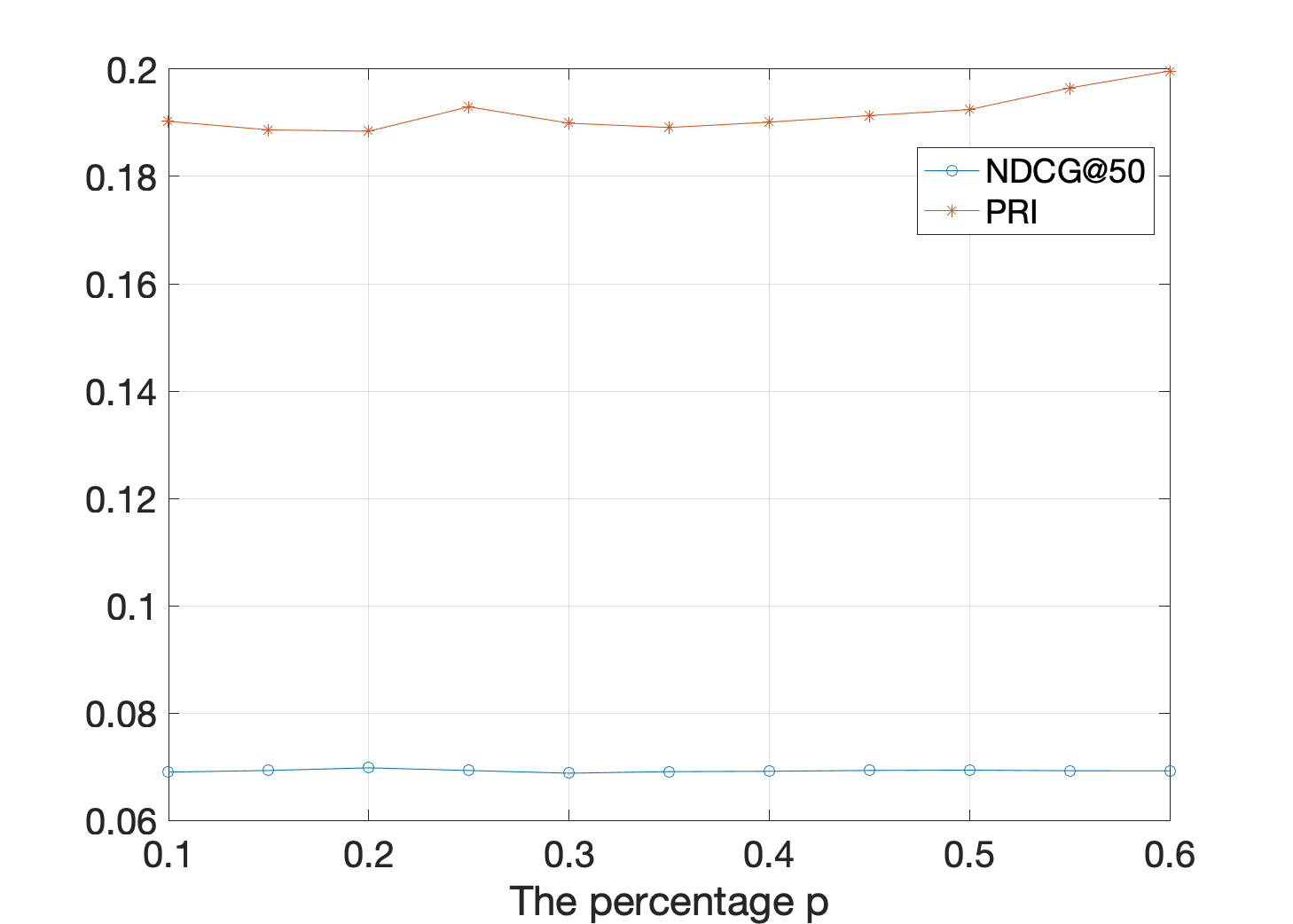}\vspace{-5pt}
		\end{minipage}
	}\hspace{-5mm}
 	\subfigure[Effects of the percentage $p$ on Beibei-L]
	{
		\begin{minipage}[b]{0.49\linewidth}
			\includegraphics[width=1\linewidth]{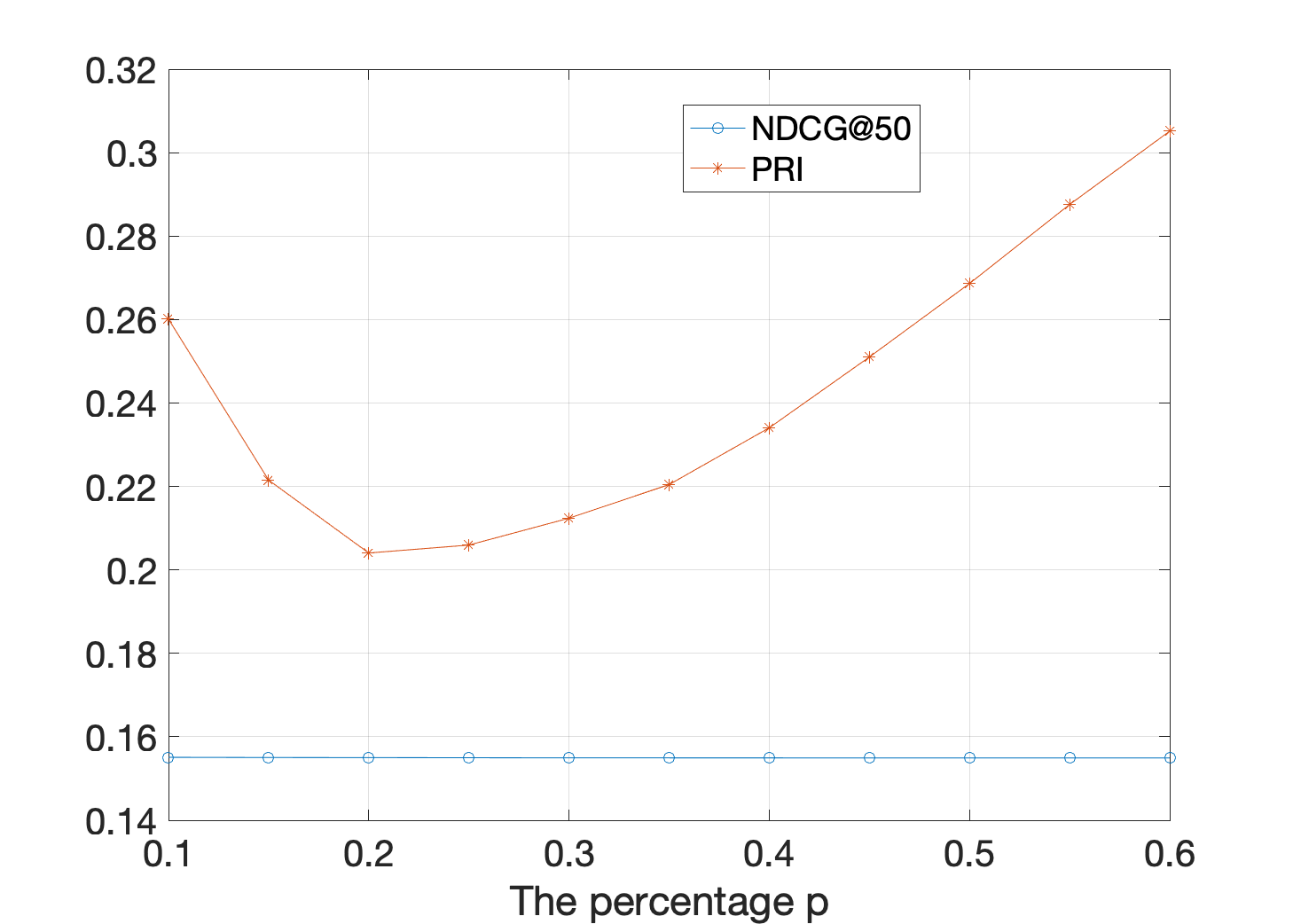}\vspace{-5pt}
		\end{minipage}
	}\hspace{-5mm}	
\end{figure}

The percentage $p$ in the popularity feature matrix $P$ demonstrates our model's sensitivity to item popularity. In Figure \ref{sensi_p}, we illustrate the recommendation accuracy (represented by $NDCG@50$) and debias performance (represented by $PRI$) of PopSI with different values of the percentage $p$ by fixing the slice rank $r$ as 200 on two chosen datasets. The larger the value of $p$, PopSI models more items to the \textit{popular} category. As can be seen, $p$ has a marginal effect on the recommendation accuracy but has an impact on the debias performance. These results are aligned with our chosen $p=0.2$, which are widely adopted with the established practices within recommender systems.

\vspace{-2mm}
\subsection{Efficiency Analysis}
Based on the aforementioned experimental results, we can confirm the effectiveness of the PopSI method on both recommendation accuracy and debias performance with a marginal tradeoff and minimal parameter tuning. Nonetheless, one still needs to consider the potential costs linked to the incorporation of multi-behavior feedback and optimization procedures. This assessment holds significant importance in evaluating the efficiency of an algorithm and its feasibility for large-scale commercial applications.
     
The time complexity of our PopSI algorithm is primarily determined by (partial) matrix singular value decompositions (SVD) on sparse matrices. 
Exact SVD of a $m \times n$ matrix has a time complexity of $O\left(\min \left\{m n^2, m^2 n\right\}\right)$, in this paper, the involved mode-$k$ unfolding matrices are assumed with low-rankness and thus we only need a few principal singular vectors/values. Besides, there are established linear algebra techniques that efficiently compute the dominant singular vectors/values by leveraging data sparsity. For example, packages like PROPACK \cite{larsen1998lanczos} can compute a rank-$r$ SVD with a cost of $O\left(\min \left\{m^2 r, n^2 r\right\}\right)$, which can be advantageous when $r \ll m, n$. And matrix-vector multiplications can be operated with running time linear in the number of non-zero elements of the matrix.
In fact, our algorithm is faster than the baseline matrix-based debiasing method MF-PC. Among the four datasets, Tmall-S and Beibei-L datasets have the smallest and largest sizes, respectively. Our method needs 10s and 1233s, respectively, on these two datasets, while MF-PC takes 20s and 18974s.
           
\end{spacing}
 
\vspace{-4mm} 
\section{Conclusion and Future Work}\label{sec-conclusion}
\begin{spacing}{2.0}
\vspace{-2mm}
\subsection{Conclusion}
Many modern recommender systems are plagued by the persistent popularity bias where algorithms prioritize already popular items disproportionately, at the cost of neglecting less popular ones that could be valuable to users. In this paper, we introduced a novel approach for building popularity-aware top-$K$ recommendation algorithm by integrating multi-behavior side information with orthogonality constraint.
Specifically, by leveraging multiple user feedback that mirrors similar user preferences and formulating it as a three-dimensional tensor, PopSI can utilize all slices to capture user preferences within the desiring slice more effectively than matrix-based approaches. Subsequently, we introduced a novel orthogonality constraint to refine the estimated item feature space, enforcing it to be invariant to item popularity by orthogonal feature projection, thereby preventing our model from the influence of popularity bias. 
Comprehensive experiments on real-world e-commerce datasets demonstrate the general improvements of the proposed method over state-of-the-art debias approaches on both recommendation accuracy and debias performance with a marginal tradeoff. 

\vspace{-2mm}
\subsection{Contribution}
In summary, this paper makes the following contributions: 
(1) We proposed a new popularity-aware recommendation framework by combining multi-behavior side information and a novel orthogonality constraint debias design together. Prevailing debias methods failed to integrate multi-behavior data, while the orthogonality constraint addresses the sensitivity of our approach to the influence of popularity bias, also preserves the expressiveness of other latent factors. 
(2) Experimental results demonstrate a significant improvement over existing baselines and debias approaches, indicating a substantial improvement in recommendation accuracy alongside enhanced debias performance with a marginal tradeoff, which is rare in existing debias methods. 
(3) Utilizing primarily matrix singular value decomposition on sparse matrices and orthogonal projections, our method exhibits low computational complexity with minimal parameter tuning and no additional training. This approach is easy to implement and finds value in product promotion for various practical scenarios.
\vspace{-2mm}
\subsection{Future Direction}  
It is noteworthy that leveraging social network connections between users can provide valuable additional side information (i.e., graph information). Understanding user behavior through these connections is crucial. Consequently, we plan to incorporate graph information to further harness the available data. 
Furthermore, there are many applications for tensors of dimension greater than three. For example, retailers might view their sales transactions over time, producing a three-dimensional tensor where time is the third dimension. Time-series data for multiple interactions may then be viewed as a four-dimensional tensor. There may be ways that the PopSI algorithm can be generalized to higher dimensions, which is worth exploring for adopting to such practical scenarios.
\end{spacing}

\section*{Acknowledgements}
    \begin{spacing}{2.0}
    \vspace{-2mm}
        This work was supported in part by the National Natural Science Foundation of China under Grant 12371513.
    \end{spacing}

\begin{spacing}{1.5}
    \bibliographystyle{elsarticle-num} 
    \bibliography{PopSI}

\begin{thebibliography}{10}
\expandafter\ifx\csname url\endcsname\relax
  \def\url#1{\texttt{#1}}\fi
\expandafter\ifx\csname urlprefix\endcsname\relax\def\urlprefix{URL }\fi
\expandafter\ifx\csname href\endcsname\relax
  \def\href#1#2{#2} \def\path#1{#1}\fi

\bibitem{chu2020position}
L.~Y. Chu, H.~Nazerzadeh, H.~Zhang, Position ranking and auctions for online marketplaces, Management Science 66~(8) (2020) 3617--3634.

\bibitem{kokkodis2023good}
M.~Kokkodis, P.~G. Ipeirotis, The good, the bad, and the unhirable: Recommending job applicants in online labor markets, Management Science (2023).

\bibitem{farias2019learning}
V.~F. Farias, A.~A. Li, Learning preferences with side information, Management Science 65~(7) (2019) 3131--3149.

\bibitem{rendle2012bpr}
S.~Rendle, C.~Freudenthaler, Z.~Gantner, L.~Schmidt-Thieme, Bpr: Bayesian personalized ranking from implicit feedback, in: Proceedings of the Twenty-Fifth Conference on Uncertainty in Artificial Intelligence, 2009, pp. 452--461.

\bibitem{he2018pseudo}
Y.~He, H.~Chen, Z.~Zhu, J.~Caverlee, Pseudo-implicit feedback for alleviating data sparsity in top-k recommendation, in: 2018 IEEE International Conference on Data Mining (ICDM), IEEE, 2018, pp. 1025--1030.

\bibitem{adomavicius2016classification}
G.~Adomavicius, J.~Zhang, Classification, ranking, and top-k stability of recommendation algorithms, INFORMS Journal on Computing 28~(1) (2016) 129--147.

\bibitem{steck2011item}
H.~Steck, Item popularity and recommendation accuracy, in: Proceedings of the fifth ACM Conference on Recommender Systems, 2011, pp. 125--132.

\bibitem{abdollahpouri2019managing}
H.~Abdollahpouri, R.~Burke, B.~Mobasher, Managing popularity bias in recommender systems with personalized re-ranking, in: Proceedings of the Thirty-Second International Florida Artificial Intelligence Research Society Conference, 2019, pp. 413--418.

\bibitem{chen2023bias}
J.~Chen, H.~Dong, X.~Wang, F.~Feng, M.~Wang, X.~He, Bias and debias in recommender system: A survey and future directions, ACM Transactions on Information Systems 41~(3) (2023) 1--39.

\bibitem{zhao2022popularity}
Z.~Zhao, J.~Chen, S.~Zhou, X.~He, X.~Cao, F.~Zhang, W.~Wu, Popularity bias is not always evil: Disentangling benign and harmful bias for recommendation, IEEE Transactions on Knowledge and Data Engineering (2022).

\bibitem{zhang2021causal}
Y.~Zhang, F.~Feng, X.~He, T.~Wei, C.~Song, G.~Ling, Y.~Zhang, Causal intervention for leveraging popularity bias in recommendation, in: Proceedings of the 44th International ACM SIGIR Conference on Research and Development in Information Retrieval, 2021, pp. 11--20.

\bibitem{zhu2021popularity}
Z.~Zhu, Y.~He, X.~Zhao, Y.~Zhang, J.~Wang, J.~Caverlee, Popularity-opportunity bias in collaborative filtering, in: Proceedings of the 14th ACM International Conference on Web Search and Data Mining, 2021, pp. 85--93.

\bibitem{brynjolfsson2006niches}
E.~Brynjolfsson, Y.~J. Hu, M.~D. Smith, From niches to riches: Anatomy of the long tail, Sloan management review 47~(4) (2006) 67--71.

\bibitem{Oestreicher2012Recommendation}
Oestreicher-Singer, Sundararajan, Recommendation networks and the long tail of electronic commerce, Mis Quarterly 36~(1) (2012) 65.

\bibitem{abdollahpouri2020connection}
H.~Abdollahpouri, M.~Mansoury, R.~Burke, B.~Mobasher, The connection between popularity bias, calibration, and fairness in recommendation, in: Proceedings of the 14th ACM Conference on Recommender Systems, 2020, pp. 726--731.

\bibitem{aridor2020deconstructing}
G.~Aridor, D.~Goncalves, S.~Sikdar, Deconstructing the filter bubble: User decision-making and recommender systems, in: Proceedings of the 14th ACM Conference on Recommender Systems, 2020, pp. 82--91.

\bibitem{mansoury2020feedback}
M.~Mansoury, H.~Abdollahpouri, M.~Pechenizkiy, B.~Mobasher, R.~Burke, Feedback loop and bias amplification in recommender systems, in: Proceedings of the 29th ACM International Conference on Information \& Knowledge Management, 2020, pp. 2145--2148.

\bibitem{adomavicius2013recommender}
G.~Adomavicius, J.~C. Bockstedt, S.~P. Curley, J.~Zhang, Do recommender systems manipulate consumer preferences? a study of anchoring effects, Information Systems Research 24~(4) (2013) 956--975.

\bibitem{canamares2018should}
R.~Ca{\~n}amares, P.~Castells, Should i follow the crowd? a probabilistic analysis of the effectiveness of popularity in recommender systems, in: The 41st International ACM SIGIR Conference on Research \& Development in Information Retrieval, 2018, pp. 415--424.

\bibitem{schnabel2016recommendations}
T.~Schnabel, A.~Swaminathan, A.~Singh, N.~Chandak, T.~Joachims, Recommendations as treatments: Debiasing learning and evaluation, in: International Conference on Machine Learning, PMLR, 2016, pp. 1670--1679.

\bibitem{wei2021model}
T.~Wei, F.~Feng, J.~Chen, Z.~Wu, J.~Yi, X.~He, Model-agnostic counterfactual reasoning for eliminating popularity bias in recommender system, in: Proceedings of the 27th ACM SIGKDD Conference on Knowledge Discovery \& Data Mining, 2021, pp. 1791--1800.

\bibitem{zheng2021disentangling}
Y.~Zheng, C.~Gao, X.~Li, X.~He, Y.~Li, D.~Jin, Disentangling user interest and conformity for recommendation with causal embedding, in: Proceedings of the Web Conference 2021, 2021, pp. 2980--2991.

\bibitem{bonner2018causal}
S.~Bonner, F.~Vasile, Causal embeddings for recommendation, in: Proceedings of the 12th ACM conference on recommender systems, 2018, pp. 104--112.

\bibitem{koren2009matrix}
Y.~Koren, R.~Bell, C.~Volinsky, Matrix factorization techniques for recommender systems, Computer 42~(8) (2009) 30--37.

\bibitem{zhao2014leveraging}
T.~Zhao, J.~McAuley, I.~King, Leveraging social connections to improve personalized ranking for collaborative filtering, in: Proceedings of the 23rd ACM International Conference on Conference on Information and Knowledge Management, 2014, pp. 261--270.

\bibitem{duan2022combining}
R.~Duan, C.~Jiang, H.~K. Jain, Combining review-based collaborative filtering and matrix factorization: A solution to rating's sparsity problem, Decision Support Systems 156 (2022) 113748.

\bibitem{guo2020consumer}
M.~Guo, X.~Liao, J.~Liu, Q.~Zhang, Consumer preference analysis: A data-driven multiple criteria approach integrating online information, Omega 96 (2020) 102074.

\bibitem{li2021personalized}
Y.~Li, R.~Wang, G.~Nan, D.~Li, M.~Li, A personalized paper recommendation method considering diverse user preferences, Decision Support Systems 146 (2021) 113546.

\bibitem{ren2024consumer}
P.~Ren, X.~Liu, W.-G. Zhang, Consumer preference analysis: Diverse preference learning with online ratings, Omega 125 (2024) 103019.

\bibitem{xv2022neutralizing}
G.~Xv, C.~Lin, H.~Li, J.~Su, W.~Ye, Y.~Chen, Neutralizing popularity bias in recommendation models, in: Proceedings of the 45th International ACM SIGIR Conference on Research and Development in Information Retrieval, 2022, pp. 2623--2628.

\bibitem{calmon2017optimized}
F.~Calmon, D.~Wei, B.~Vinzamuri, K.~Natesan~Ramamurthy, K.~R. Varshney, Optimized pre-processing for discrimination prevention, Advances in Neural Information Processing Systems 30 (2017).

\bibitem{chen2022learning}
J.~Chen, D.~Lian, B.~Jin, K.~Zheng, E.~Chen, Learning recommenders for implicit feedback with importance resampling, in: Proceedings of the ACM Web Conference 2022, 2022, pp. 1997--2005.

\bibitem{rhee2022countering}
W.~Rhee, S.~M. Cho, B.~Suh, Countering popularity bias by regularizing score differences, in: Proceedings of the 16th ACM Conference on Recommender Systems, 2022, pp. 145--155.

\bibitem{abdollahpouri2021user}
H.~Abdollahpouri, M.~Mansoury, R.~Burke, B.~Mobasher, E.~Malthouse, User-centered evaluation of popularity bias in recommender systems, in: Proceedings of the 29th ACM Conference on User Modeling, Adaptation and Personalization, 2021, pp. 119--129.

\bibitem{bertsimas2023tensor}
D.~Bertsimas, C.~Pawlowski, Tensor completion with noisy side information, Machine Learning 112~(10) (2023) 3945--3976.

\bibitem{ding2020improving}
J.~Ding, G.~Yu, Y.~Li, X.~He, D.~Jin, Improving implicit recommender systems with auxiliary data, ACM Transactions on Information Systems (TOIS) 38~(1) (2020) 1--27.

\bibitem{liu2017personalized}
J.~Liu, C.~Shi, B.~Hu, S.~Liu, P.~S. Yu, Personalized ranking recommendation via integrating multiple feedbacks, in: Advances in Knowledge Discovery and Data Mining: 21st Pacific-Asia Conference, PAKDD 2017, Jeju, South Korea, May 23-26, 2017, Proceedings, Part II 21, Springer, 2017, pp. 131--143.

\bibitem{chen2020efficient}
C.~Chen, M.~Zhang, Y.~Zhang, W.~Ma, Y.~Liu, S.~Ma, Efficient heterogeneous collaborative filtering without negative sampling for recommendation, in: Proceedings of the AAAI conference on artificial intelligence, Vol.~34, 2020, pp. 19--26.

\bibitem{meng2023hierarchical}
C.~Meng, H.~Zhang, W.~Guo, H.~Guo, H.~Liu, Y.~Zhang, H.~Zheng, R.~Tang, X.~Li, R.~Zhang, Hierarchical projection enhanced multi-behavior recommendation, in: Proceedings of the 29th ACM SIGKDD Conference on Knowledge Discovery and Data Mining, 2023, pp. 4649--4660.

\bibitem{he2017neural}
X.~He, L.~Liao, H.~Zhang, L.~Nie, X.~Hu, T.-S. Chua, Neural collaborative filtering, in: Proceedings of the 26th International Conference on World Wide Web, 2017, pp. 173--182.

\bibitem{rao2015collaborative}
N.~Rao, H.-F. Yu, P.~K. Ravikumar, I.~S. Dhillon, Collaborative filtering with graph information: Consistency and scalable methods, Advances in Neural Information Processing Systems 28 (2015).

\bibitem{hu2008collaborative}
Y.~Hu, Y.~Koren, C.~Volinsky, Collaborative filtering for implicit feedback datasets, in: 2008 Eighth IEEE International Conference on Data Mining, Ieee, 2008, pp. 263--272.

\bibitem{he2016fast}
X.~He, H.~Zhang, M.-Y. Kan, T.-S. Chua, Fast matrix factorization for online recommendation with implicit feedback, in: Proceedings of the 39th International ACM SIGIR conference on Research and Development in Information Retrieval, 2016, pp. 549--558.

\bibitem{chen2023revisiting}
C.~Chen, W.~Ma, M.~Zhang, C.~Wang, Y.~Liu, S.~Ma, Revisiting negative sampling vs. non-sampling in implicit recommendation, ACM Transactions on Information Systems 41~(1) (2023) 1--25.

\bibitem{kolda2009tensor}
T.~G. Kolda, B.~W. Bader, Tensor decompositions and applications, SIAM review 51~(3) (2009) 455--500.

\bibitem{nickel2011three}
M.~Nickel, V.~Tresp, H.-P. Kriegel, et~al., A three-way model for collective learning on multi-relational data., in: Proceedings of the 28th International Conference on Machine Learning, Vol.~11, 2011.

\bibitem{abdollahpouri2017controlling}
H.~Abdollahpouri, R.~Burke, B.~Mobasher, Controlling popularity bias in learning-to-rank recommendation, in: Proceedings of the eleventh ACM conference on recommender systems, 2017, pp. 42--46.

\bibitem{wei2022contrastive}
W.~Wei, C.~Huang, L.~Xia, Y.~Xu, J.~Zhao, D.~Yin, Contrastive meta learning with behavior multiplicity for recommendation, in: Proceedings of the fifteenth ACM international conference on web search and data mining, 2022, pp. 1120--1128.

\bibitem{xia2021graph}
L.~Xia, Y.~Xu, C.~Huang, P.~Dai, L.~Bo, Graph meta network for multi-behavior recommendation, in: Proceedings of the 44th international ACM SIGIR conference on research and development in information retrieval, 2021, pp. 757--766.

\bibitem{steck2019collaborative}
H.~Steck, Collaborative filtering via high-dimensional regression, CoRR, abs/1904.13033 (2019).

\bibitem{boratto2021connecting}
L.~Boratto, G.~Fenu, M.~Marras, Connecting user and item perspectives in popularity debiasing for collaborative recommendation, Information Processing \& Management 58~(1) (2021) 102387.

\bibitem{larsen1998lanczos}
R.~M. Larsen, Lanczos bidiagonalization with partial reorthogonalization, DAIMI Report Series~(537) (1998).

\end{thebibliography}
\end{spacing}

\end{document}